\documentclass[10pt,a4paper,twocolumn,hidelinks]{article}

\usepackage{latexsym}      
\usepackage{hyperref}
\hypersetup{
 colorlinks=false,
 citecolor=blue,
 linkcolor=blue,
 urlcolor=blue,
 pdfpagemode=UseNone,
 pdfstartview=FitH}
\usepackage{graphicx}      
\usepackage{amsmath}       
\usepackage[affil-it]{authblk} 			
\usepackage[margin=2cm]{geometry} 

\usepackage{sectsty}

\makeatletter
\newcommand{\Capitalize}[1]{%
  \edef\@tempa{\expandafter\@gobble\string#1}%
  \edef\@tempb{\expandafter\@car\@tempa\@nil}%
  \edef\@tempa{\expandafter\@cdr\@tempa\@nil}%
  \uppercase\expandafter{\expandafter\def\expandafter\@tempb\expandafter{\@tempb}}%
  \@namedef{\@tempb\@tempa}{\expandafter\MakeUppercase\expandafter{#1}}}
\makeatother

\makeatletter
\newbox\abstract@box
\renewenvironment{abstract}
  {\global\setbox\abstract@box=\vbox\bgroup
     \hsize=\textwidth\linewidth=\textwidth
    \small
    \begin{center}%
    {\bfseries \abstractname\vspace{-.5em}\vspace{\z@}}%
    \end{center}%
    \quotation}
  {\endquotation\egroup}
\expandafter\def\expandafter\@maketitle\expandafter{\@maketitle
  \ifvoid\abstract@box\else\unvbox\abstract@box\if@twocolumn\vskip1.5em\fi\fi}
\makeatother
\providecommand{\keywords}[1]{\noindent \mdseries{{Keywords:}} #1}
\providecommand{\DOI}[1]{\vspace{1\baselineskip}}

\makeatletter
\renewcommand\section{\@startsection
   {section}{1}{0pt}%
   {-\baselineskip}%
   {0.3\baselineskip}%
   {\normalfont\bfseries}}%
\makeatother
\renewcommand\thesection{\Roman{section}}

\makeatletter
\renewcommand\subsection{\@startsection
   {subsection}{1}{0pt}%
   {-\baselineskip}%
   {0.3\baselineskip}%
   {\normalfont\bfseries}}%
\makeatother
\renewcommand{\thesubsection}{\thesection.\arabic{subsection}}

\usepackage{booktabs}
\usepackage{dcolumn}
\newcolumntype{d}[1]{D{.}{.}{#1}}

\usepackage[labelfont=bf]{caption}
\usepackage{fancyhdr}
\fancypagestyle{titlestyle}
{

\fancyhf[c]{ }
\fancyhf[r]{ }
\fancyfoot[c]{1}
}

\usepackage{bm}

\usepackage{textcomp}
\usepackage{graphicx}
\usepackage{dcolumn}
\usepackage{bm}
\usepackage{tikz}
\usetikzlibrary{shapes.geometric, arrows}
\usetikzlibrary{positioning}
\usetikzlibrary{shapes,arrows}
\usetikzlibrary{intersections}
\usepackage{csvsimple}
\usepackage{changepage}
\usepackage[tbtags]{mathtools}
\usepackage{siunitx}
\usepackage{csquotes}
\usepackage{enumerate}
\usepackage{enumitem}
\usepackage{float}
\usepackage[label font=bf]{subfig}
\usepackage{amsmath}
\usepackage{wasysym}

 \usepackage[
 	nonumberlist, 				
 	acronym,
 	nomain,
 	nopostdot,
 ]{glossaries}
 \usepackage{glossary-superragged}
\usepackage{xcolor}

\usepackage{pgfplots}
\usepackage{tikz}
\usetikzlibrary{arrows}
\usepackage{amsmath}
\pgfplotsset{compat=newest}
\usepgfplotslibrary{fillbetween}
\usetikzlibrary{calc}
\def\centerarc[#1](#2)(#3:#4:#5)
{ \draw[#1] ($(#2)+({#5*cos(#3)},{#5*sin(#3)})$) arc (#3:#4:#5); }
\usetikzlibrary{external}
\usepackage{blindtext}
\usepackage{amssymb}

\usepackage{nicefrac}

\usepackage{multirow}
\usepackage{tabularx}
\usepackage{booktabs}
\usepackage{array}
\usepackage{longtable}
\newcolumntype{L}[1]{>{\raggedright\let\newline\\\arraybackslash\hspace{0pt}}m{#1}}
\newcolumntype{C}[1]{>{\centering\let\newline\\\arraybackslash\hspace{0pt}}m{#1}}
\newcolumntype{R}[1]{>{\raggedleft\let\newline\\\arraybackslash\hspace{0pt}}m{#1}}

\usepackage[numbers,square,sort&compress]{natbib}
\usepackage[normalem]{ulem}
\usepackage{microtype}

\newcommand\cori[1]{%
  {\textcolor{black}{#1}}%
}
\definecolor{correccolortwo}{rgb}{0.85, 0.0, 0.1}
\newcommand\corii[1]{%
  {\textcolor{black}{#1}}%
}
\newcommand\coriii[1]{%
  {\textcolor{black}{#1}}%
}

\definecolor{deleted}{rgb}{0.66, 0.66, 0.66}

\usepackage{titlesec}
\titleformat{\section}
  {\bf\sffamily}
  {\thesection. }
  {5pt}
  {\MakeUppercase}
\renewcommand{\thesection}{\Roman{section}} 

\titleformat{\subsection}
  {\bf\sffamily}
  {\thesubsection. }
  {5pt}{}

\newacronym{3d}{3D}{three dimensional}
\newacronym{dc}{DC}{direct-current}
\newacronym[plural=PFCs,firstplural=parabolic flight campaigns (PFCs)]{pfc}{PFC}{parabolic flight campaign}
\newacronym{fft}{FFT}{Fast Fourrier Transform}
\newacronym{cad}{CAD}{Computer Assisted Design}
\newacronym{ptfe}{PTFE}{polytetrafluoroethylene}
\newacronym{ps}{PS}{polystyrene}
\newacronym{esa}{ESA}{European Space Agency}
\newacronym{si}{SI}{International System of Units, abbreviated from French \textit{Syst\`{e}me International (d'unit\'{e}s)}}
\newacronym{dlr}{DLR}{German Aerospace Center, abbreviated from German \textit{Deutsches Zentrum f\"{u} Luft- und Raumfahrt e.V.}}
\newacronym{daad}{DAAD}{German Academic Exchange Service, abbreviated from German \textit{Deutscher Akademischer Austauschdienst}}
\newacronym{fsm}{FSM}{finite-state machine}
\newacronym{ir}{IR}{infrared}
\newacronym{pcbs}{PCBs}{Printed Circuit Boards}
\newacronym{pcb}{PCB}{Printed Circuit Board}
\newacronym{mcr}{MCR}{Modular Compact Rheometer}
\newacronym{sff}{SFF}{Solid Freeform Fabrication}
\newacronym{uv}{UV}{ultraviolet}
\newacronym{abs}{ABS}{acrylonitrile butadiene styrene}
\newacronym{hpde}{HPDE}{high density polyethylene}
\newacronym{pei}{PEI}{polyetherimide}
\newacronym{bff}{BFF}{BioFabrication Facility}
\newacronym{lens}{LENS}{Laser Engineered Net Shaping}
\newacronym{cnc}{CNC}{Computer Numerical Control}
\newacronym{ebf3}{EBF$^3$}{Electron Beam Free-Form Fabrication}
\newacronym{leo}{LEO}{Low Earth Orbit}
\newacronym{pc}{PC}{polycarbonate}
\newacronym{2d}{2D}{two dimensional}
\newacronym{ft4}{FT4}{Freeman Technology 4 Powder Rheometer}
\newacronym{dsc}{DSC}{Differential Scanning Calorimetry}
\newacronym{pmma}{PMMA}{polymethylmethacrylate}
\newacronym{1g}{gnd}{on-ground}
\newacronym{mug}{$\textrm{$\mu$-g}$}{microgravity}
\newacronym{npi}{NPI}{Network Partnering Initiative}
\newacronym{ecsat}{ECSAT}{European Centre for Space Applications and Telecommunications}
\newacronym{eac}{EAC}{European Astronaut Centre}
\newacronym{estec}{ESTEC}{European Space Research and Technology Centre}
\newacronym{fps}{fps}{frames per second}
\newacronym{pdf}{pdf}{probability density function}
\newacronym{al}{Al}{aluminium}
\newacronym{ss}{\textit{SS}}{\textit{Smooth Surface}}
\newacronym{rs}{\textit{RS}}{\textit{Rough Surface}}
\newacronym{isru}{ISRU}{\textit{in-situ} resource utilization}
\newacronym{rlp}{rlp}{random loose packing}
\newacronym[plural=rcps,firstplural=random close packings (rcps)]{rcp}{rcp}{random close packing}
\newacronym{dfg}{DFG}{German Physics Foundation}
\newacronym{sgr}{SGR}{soft glassy rheology}

\begin{document}

\title{\textbf{Granular piston-probing in microgravity:\\
powder compression, from densification to jamming}}

\author[1,2]{Olfa D'Angelo\thanks{Corresponding author: olfa.dangelo@mail.com}}
\author[3]{Anabelle Horb\thanks{Current address: Omnidea, Lda.~-- Polypark, N\`{u}cleo Empresarial da Arruda dos Vinhos, Estrada da Quinta de Matos 4, 2630-179 Arruda~dos~Vinhos, Portugal.}}
\author[3]{Aidan Cowley}
\author[1,4]{Matthias Sperl}
\author[4,1]{W.~Till Kranz}

\affil[1]{Institut f\"ur Materialphysik im Weltraum, Deutsches Zentrum f\"ur Luft- und Raumfahrt (DLR), 51170 K\"oln, Germany.}
\affil[2]{Institute for Multiscale Simulation, Universit\"{a}t Erlangen-N\"{u}rnberg, Cauerstra\ss{}e 3, 91058 Erlangen, Germany.}
\affil[3]{European Astronaut Centre (EAC), European Space Agency (ESA), 51170 K\"{o}ln, Germany.}
\affil[4]{Institut f\"ur Theoretische Physik, Universit\"at zu K\"oln, 50937 K\"oln, Germany.}

\date{\today}

\begin{abstract}
\noindent
The macroscopic response of granular solids is determined by
the microscopic fabric of force chains, which, in turn, is intimately
linked to the history of the solid. 
To query the influence of gravity on powder flow-behavior, 
a granular material is subjected to compression by a piston in a closed container,
on-ground and in microgravity (on parabolic flights).
Results show that piston-probing densifies the packing,
eventually leading to jamming of the material compressed by the piston,
regardless of the gravitational environment. 
The onset of jamming is found to appear at lower packing fraction
in microgravity
\corii{($\varphi^{\textrm{$\mu$-g}}_J = 0.567 \pm 0.014$)}
than on-ground
\corii{($\varphi^{\text{gnd}}_J = 0.579 \pm 0.014 $)}.
We interpret these findings 
as the manifestation of a granular fabric
altered by the gravitational force field:
in absence of a secondary load (due to gravitational acceleration) to stimulate reorganization 
in a different direction to the major compression stress,
the particles' configuration becomes stable at lower density,
as the particles have no external drive
to promote reorganization into a denser packing.
This is coupled with a change in interparticular force balance 
which takes place under low gravity, as cohesive interactions become predominant.
We propose
a combination of microscopic and continuum arguments
to rationalize our results.
\linebreak


\end{abstract}

\begingroup
\sffamily
\maketitle
\endgroup

\thispagestyle{titlestyle}

\section*{Introduction}\label{sec:intro}
As humans make their way into space
and plan to travel 
to sand-covered celestial bodies,
the influence of (reduced) gravity on granular phenomena becomes
a matter of concern~\cite{Wilkinson2005},
and new strategies are sought
to handle granular materials efficiently
in various gravitational environments~\cite{Karapiperis2020, Antony2021, dangelo2021, Mo2021}.
While the action of gravity 
is intuitively expected to
weaken as objects become smaller,
the collective behavior of small particles composing granular systems is
significantly modified 
by the  
gravitational field,
challenging our expectation of powder response outside of Earth gravity.
Granular media are by definition composed of grains
\emph{big enough} not to be subject to thermal motion~\cite{deGennes1999, Frenkel2002} but \emph{small enough}
for continuous material properties to emerge on the mesoscopic scale. 
Such grains are, therefore, generally
sufficiently large for gravitational forces on each of them not to be negligible in comparison to the other forces in presence.
The flow behavior of granular matter can thus change dramatically with a change in gravitational acceleration,
such as that 
affecting a body on a foreign celestial surface (e.g.~Moon, Mars, or an asteroid), or subjected to microgravity conditions.

The aim of this article is to
investigate  
the densification of a granular packing
in two scenarios:
\gls{1g}, where Earth's gravity may accelerate particles relative to the fixed container, and 
in \gls{mug},
on parabolic flights, where gravity cannot induce such relative acceleration. Note, that the term \emph{microgravity} is widely used, although \emph{free-fall} would be more precise \cite{Karmali2008}.

We propose a densification mechanism in the form of a piston 
rising quasi-statically through a confined powder bed (see
Fig.~\ref{fig:piston_probing_setup}), thereby compressing the material placed above.
The piston is of diameter inferior to that of the container. 
Compression is continued until reaching a jammed state above the piston,
hence probing
the jamming point variations 
between 
the two environments.
The jammed state is characterized by 
the appearance of a finite elastic modulus of the granular packing,
which drives the piston to a stop.
Such piston mechanism 
could be used,
on-ground or in space,
to densify
shallow granular beds
and probe the percolation of the force network, possibly
before submitting the material to further processing.

In presence of gravity, granular flows can be caused by their
potential energy in gravity-driven flows, providing a granular
time-scale in the form of the time needed for a grain to fall by its
diameter.  In microgravity, particles do not fall towards the
minimum-height position available; instead, they cluster into floating
aggregates~\cite{Blum2000, Blum2006, Weidling2012, Love2014}, as the
interparticle cohesive forces become the predominant interaction.  If
no force field sediments the packing, particles or clusters are not
necessarily in contact with their closest
neighbors~\cite{Evesque2004}.  Force
chains~\cite{Drescher1972,Liu1995}, enabling force transmission
through lasting contacts in a dense granular medium, cannot appear
unless a contact network is established. Yet, such contact network
can also trigger flow arrest on the macroscopic scale, if it
percolates into a force network spanning the system, transmitting
force to its boundaries akin to a solid~\cite{Liu2001, Liu1998,
Brown2014}.

The {jamming transition}~\cite{Majmudar2007, Behringer2018}
describes the changeover between the flowing state, in which the 
granular fabric deforms plastically,
and the jammed state, where
a mechanically stable contact network is formed,
which can withstand a certain amount of stress without reorganizing.
In other words,
the granular assembly as a whole 
exhibits a finite elastic modulus.
If unpredicted,
the appearance of jammed regions can be 
detrimental to
industrial processes 
involving handling and transport of granular materials,
as it can draw entire production processes to a halt.
In sheared granular media, an anisotropic {shear-jammed} phase
can block powder flow 
mono-directionally~\cite{Cates1999,Bi2011,Wang2013,Grob2014}.
It was suggested that for frictionless particles, this transition
already occurs at \gls{rlp}~\cite{Onoda1990, Metayer2011}, 
while a granulate in the state of \acrlong{rcp} would
be isotropically jammed~\cite{Song2008, Torquato2000}.
However, the existence of a
\emph{fixed} jamming packing fraction $\varphi_\text{J}$ ({jamming point}) has now been challenged~\cite{Kumar2016, Chaudhuri2010}.

Besides evident dependence on particles'
shape~\cite{Torquato2010,VanHecke2009}, (poly)dispersity~\cite{Hermes2010} or mechanical properties~\cite{VanHecke2009},
experimental and numerical studies have shown the 
history-dependence 
of $\varphi_\text{J}$,
with variations, even for a single granular material, 
related to
the preparation protocol~\cite{Torquato2010, Kumar2016, Chaudhuri2010, Ciamarra2010, Otsuki2012}
or compression rate of the granular bed leading to jamming
(namely, $\varphi_\text{J}$ decreases with increasing compression rate~\cite{Hermes2010, Baranau2014}).
Interparticular friction was shown to decrease the jamming
packing by decreasing particles' mobility in a dense
packing~\cite{VanHecke2009, Ciamarra2011, Mari2014, Bi2011}.

It has been reported that when
gravitational forces are negligible, 
the density of a granulate in \acrlong{rlp} 
decreases from
$\varphi^{\text{gnd}}_{\text{rlp}}=0.60$ to
$\varphi^{\textrm{$\mu$-g}}_{\text{rlp}}=0.55$~\cite{Onoda1990, Jerkins2008, Silbert2010, Farrell2010, Noirhomme2017}
(for a granular material composed of hard monodispersed spherical particles with low or no friction \cori{in three dimensions}).
However, to the best of our knowledge, the
impact of gravitational acceleration -- and particularly of very low gravity $\sim \textrm{$\mu$-g}$ -- on the packing fraction at the jamming transition 
has hitherto not been studied experimentally.

Before reporting on the actual measurements, let us make a naive
prediction. 
We consider a piston placed at the bottom of 
a container filled with granular material at a density $\varphi \approx 0.55$.
As in our setup, the piston has a diameter $d = \nicefrac{2}{3} D$, where $D$ is the container diameter.
When the piston rises
by a small height $\Delta h$,
it displaces the volume $\pi d^2\Delta h/4$ of powder above the piston.
This may either be achieved by compacting the powder to 
a higher packing fraction above the piston,
or by pushing
powder beneath the piston.
\Acrlong{1g}, the material underneath the
piston will settle in an annular heap at its angle of repose
$\alpha$. At small piston displacements $h$, the volume of that heap
grows more slowly by a factor $2h / d\tan\alpha \ll 1$ compared to the
displaced material on the top. Consequently, we expect the piston to stop
shortly after the material is fully compacted. Under \acrlong{mug}, however,
the displaced material does not have to settle in a
heap; it may instead
fill the entire volume freed below the piston. Consequently, one
may expect the piston to be able to rise indefinitely
without being stopped by a jammed phase.

\section*{Results}\label{sec:results}

\subsection*{Piston-probing}

Piston-probing consists of applying a finite amount of normal stress
to a granular material in a confined space, by raising a piston inside a
container filled by this material.
A lateral space is left on the sides of the piston's platform,
allowing the granulate to yield and flow downwards, around the platform.
If applicable, the gravity vector points downward, against the rising motion of the piston.

\begin{figure}[h!]
\includegraphics[width=\linewidth]{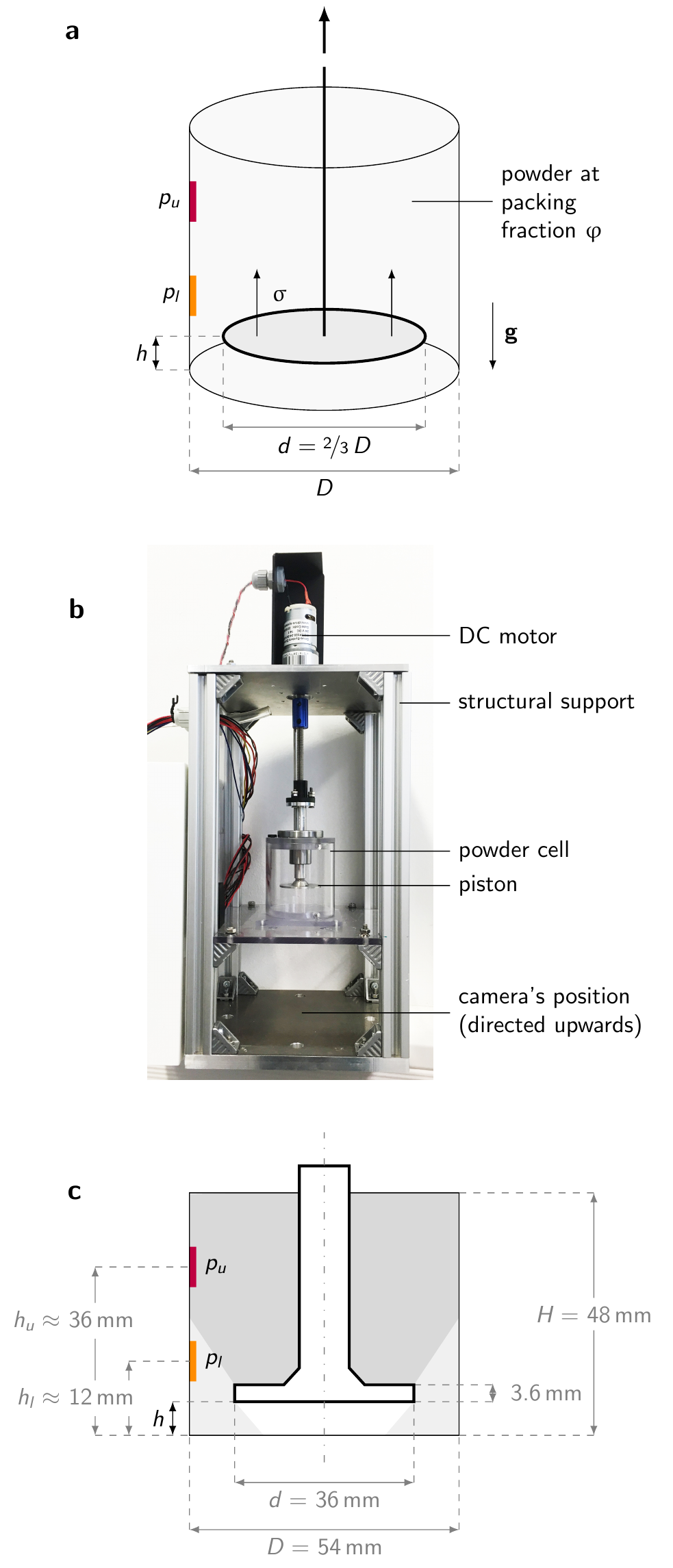}
    \caption{\label{fig:piston_probing_setup}
Piston-probing setup.
(a) Piston-probing principle: a piston rises incrementally in a closed container, filled with powder at packing fraction $\varphi$, applying a normal stress $\Vec{\sigma}$ on the granular packing.
The pressure $p$ transmitted through the granular material is recorded by two pressure sensors placed on the lateral walls of the container (respectively recording $p_u$ and $p_l$, \textit{upper} and \textit{lower} pressures).
The direction of the gravitational acceleration $\Vec{g}$ is indicated by a downwards pointing arrow.
(b)~Experimental setup, showing the container (here empty) in its support structure (photograph taken by the authors).
(c)~Schematic including container's and piston's dimensions (in grey), and variables
recorded (in black).
}
\end{figure}

The experimental setup is presented in
Fig.~\ref{fig:piston_probing_setup}.
It consists of a transparent
\acrshort{pmma} tube of diameter $D=\SI{54}{\mm}$ and height
$H=\SI{48}{\mm}$, inside which a piston of diameter $d=\SI{36}{\mm}$
can rise incrementally. 
The piston is moved by a \gls{dc} motor, driving a traveling-nut
linear translation stage that allows 
the piston to rise vertically without rotating.
The translation stage can be placed in one of two
modes by the user: (i) a stationary mode, where the piston is actively
stabilized at its current position; and (ii) a traveling mode where
the piston is actively pulling upwards. The \gls{dc} motor is
regulated through a velocity-controlled closed control loop that keeps
the velocity at its set value until the resistance of the packing
reaches a significant fraction of the motor's maximum torque,
at which point the
piston slows down and eventually stops. 
The maximum torque was chosen such, that the experiment probes
the elastic modulus of the jammed packing, yet only minimally
deforms the particles which are much stiffer (see Methods).
An optical measurement of the piston height $h$ provides an indirect
assessment of its position, thus of its {true} rising speed.

Pressure sensors are placed inside the tube, on the lateral walls, and
record the pressure evolution as the piston compresses the granular
sample. One sensor is placed on the upper part of the container
($h_u \approx $~\SI{36}{\mm}), and another one on the bottom part
($h_l \approx $~\SI{12}{\mm}).  The sensitive area of the pressure
sensors measures \SI{200}{\square\mm}, which represents approx.~30~000
particles.  The pressure sensors are calibrated in-house \cori{(see
  Methods for details)}, so the
pressure recorded corresponds to the pressure difference
due to the powder contained in the cell.
A camera below the
transparent container 
allows to observe inside the setup
from below.
At the beginning of the experiment, the piston is generally at its lowest
position, i.e.~the platform touches the bottom of the closed tube; it
is then programmed to rise step-by-step, each step being activated
by the user during approximately \SI{20}{\s}.
In some experiments the
platform is placed in the middle of the box, at its median position in
height (\textit{cf.}~results shown in Fig.~\ref{fig:qualitative}(a) and (b)). 
Hardware technical specifications are given
in Table~\ref{tab:piston_probing_tab},
and a list of all experiments reported is provided in Supplementary Material
(Supplementary Table~1).

\begin{table}
\caption{\label{tab:piston_probing_tab}References for hardware used on the piston-probing experimental setup.}
    \centering
    \small
    \begin{tabular}{l l}
        \toprule
        \textbf{Hardware} & \textbf{Model}  \\
        \midrule 
         \acrshort{dc} motor & DSMP~320-24-0189-BF \\
         			& (max.~torque \SI{0.69}{\newton\meter}) \\
         current sensor & Pololu ACS7174 \\
         pressure sensor  & SingleTact capacitive pressure sensor \\
        camera & YI 4K Plus Action Camera \\ 
        			& (resolution $3840 \times 2160$ pixels) \\
        \bottomrule 
    \end{tabular}
\end{table}

The granular material used to fill the container is a monodispersed \gls{ps}
powder with spherical particles of diameter \SI{80}{\micro\meter} 
(see Methods 
for further details).  
Inside the closed container, the mass of
granular material inserted determines an overall packing fraction of
$\varphi_0 \approx 0.55$.
At the beginning of the experiment (that is, until reaching a jammed state),
the packing fraction is assumed to be sufficiently homogeneous
to be calculated from this initial packing fraction $\varphi_0$
and the piston's position in height, $h$. 
Therefore, the 
packing fractions given throughout this article, and in particular the packing fractions at the jamming transition $\varphi_{\text{J}}$, are \emph{global} packing fractions;
the local packing fractions are not determined experimentally.
A calculation of formal uncertainty is provided in the Methods section.

Let us consider the most common case, where the piston is initially
at the bottom of the container ($h=0$).
The piston diameter being two thirds of
the container,
a lateral space is left around the piston, 
through which the granular material can flow
downward to reach the bottom of the container (below the piston). 
The dimensions of the
piston are chosen such, that the upward flow imposed by the piston
rise could be accommodated by a downward flow through the gap
around the piston. Maximum speed and force of the piston were chosen
to give the packing ample time to relax, yet allow for significant
piston motion within the limited time intervals of \acrlong{mug},
and to limit particle deformations in the jammed state.

During the initial instants of the experiment, 
the powder distribution
under the piston can be observed through a transparent bottom
plate. At all stages of the experiment, the piston position $h$ and the
pressure on the outer walls at two heights, $p_u$ and $p_l$ (respectively \textit{upper} and \textit{lower}), are recorded.

This experiment flew on the 31\textsuperscript{st} \acrshort{dlr}
\gls{pfc} in March 2018.
\Glspl{pfc} provide periods of \acrlong{mug} ($\sim 10^{-3} g$) of \SI{22}{\s}
as the plane (and its content) is in free-fall, alternating with
hypergravity ($\approx 1.8\,g$) phases (as the plane rises and swoops).
During the free-fall period, in the frame of the plane, the
directional force of gravity is replaced by approximately isotropic,
residual accelerations (termed $g$-jitter) on the level of $\sim 0.05\,g$.
Here $g=\SI{9.81}{\metre\per\second\squared}$ denotes the
standard acceleration of gravity.
For an overview of
parabolic flights as a gravity-related experimental platform, the reader
is referred to Ref.~\cite{Karmali2008}.

Experiments conducted on parabolic flights are 
reproduced on-ground,
using the same experimental setup
in the laboratory.
It should be noted that the flight time up to the first parabola, as
well as the first hypergravity phase might contribute to create an
initial state in \acrlong{mug} that differs from their \acrlong{1g} counterpart.

A \gls{2d} toy model 
reproducing a vertical section of our setup
with photoelastic particles~\cite{Zadeh2019, Daniels2017} is used to help visualizing 
the dependence of
the force chains 
network on a secondary force field acting downwards
(see Supplementary Material, Supplementary Figure~1).

\subsection*{Time evolution of pressure and density}

\begin{figure}
  \centering
\includegraphics[width=\linewidth]{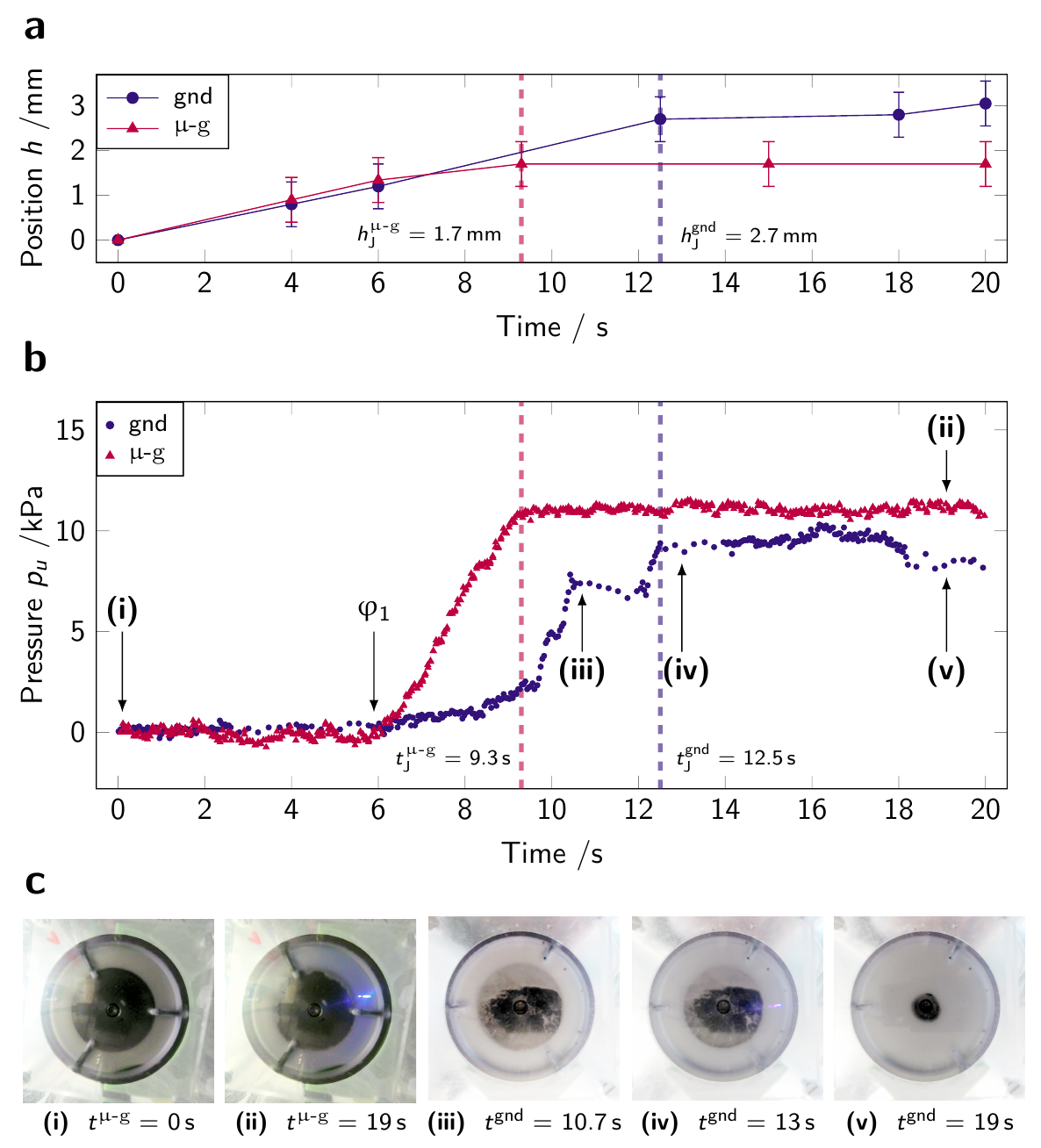}
  \caption{\label{fig:piston_probing_densification} Time evolution of piston position and pressure in powder cell.
(a)~Piston position in height as a function of time. Error bars represent the
experimental precision in optical measurement. 
(b)~Pressure recorded by the top sensor during the first step up of
the piston, \acrlong{1g} (blue disks, \gls{1g}) and in
\acrlong{mug} (red triangles, \gls{mug}). The 
jamming times $t_{\text{J}}$ are marked by vertical dashed lines, as well as $\varphi_1$ (see body of the text) and five points of interest that
correspond to panel (c), images (i) to (v). (i) and (ii) are images from the
bottom of the experiment cell looking up in \acrlong{mug}; (iii),
(iv) and (v) \acrlong{1g}. The piston is black and the
\gls{ps} powder, white. The clearly visible piston at all
times in \acrlong{mug} (i, ii, iii) indicates the absence of powder flow below the piston,
different from the behavior \acrlong{1g} (iv, v).}
\end{figure}

Pressure data from the first step up of the piston, measured on the upper half
of the container, \acrfull{1g} and in \acrfull{mug}, is presented in
Fig.~\ref{fig:piston_probing_densification}.
For all \acrlong{mug} experiments, $t = \SI{0}{\s}$ corresponds to the start of the
\acrlong{mug} period. For experiments conducted \acrlong{1g}, 
it corresponds to
the start of the piston rise.

\begin{figure}
  \centering
  \includegraphics[width=\linewidth]{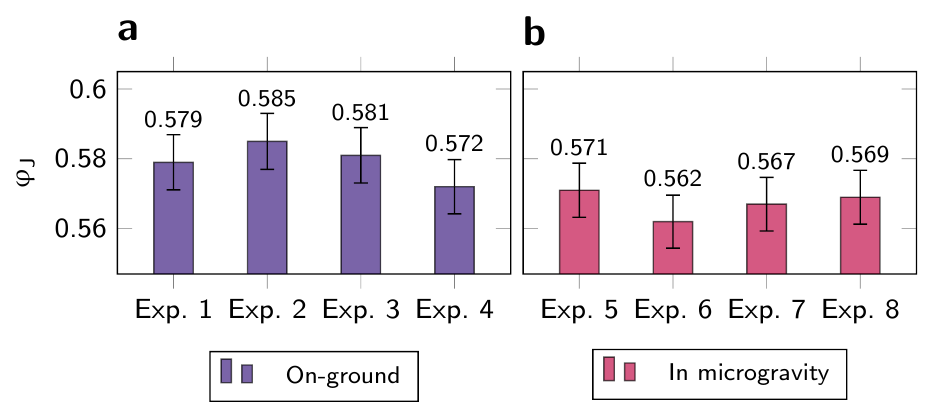}
  \caption{\label{fig:critical_height} Packing fraction at
    jamming $\varphi_{\text{J}}$, estimated from the height at which the
    piston is driven to a halt by the resistance of the packing. Experiments are conducted in the same setup,
(a)~\acrfull{1g} and (b)~in \acrfull{mug}. Error bars represent the formal accuracy (see Methods section for details on the calculation).}
\end{figure}

At first and until approximately \SI{6}{\s}, the piston is rising (Fig.~\ref{fig:piston_probing_densification}(a)) but
the pressure does not undergo any change (Fig.~\ref{fig:piston_probing_densification}(b)). In this initial phase, the
powder is compressed from the initial packing fraction $\varphi_0$ to a denser
configuration, $\varphi_1 > \varphi_0$. For both experiments,
only then does the pressure start increasing. A second characteristic
density, $\varphi_{\text{J}}$, is reached when the piston is brought to a halt
and the powder resists the maximum stress imposed on the sample. By
recording the piston's position at this instant, global estimates of
$\varphi_J$ have been obtained in four repetitions of the experiment,
each \acrlong{1g} and under \acrlong{mug} conditions;
resulting $\varphi_{\text{J}}$ for each repetition are presented in
Fig.~\ref{fig:critical_height}. Note that the jamming density is
systematically lower in \acrlong{mug} compared to ground.
Statistical significance test
shows that albeit the small number of repetitions available,
this finding is statistically significant (see Methods section).

\begin{figure*}[h]
  \centering
\includegraphics[width=\linewidth]{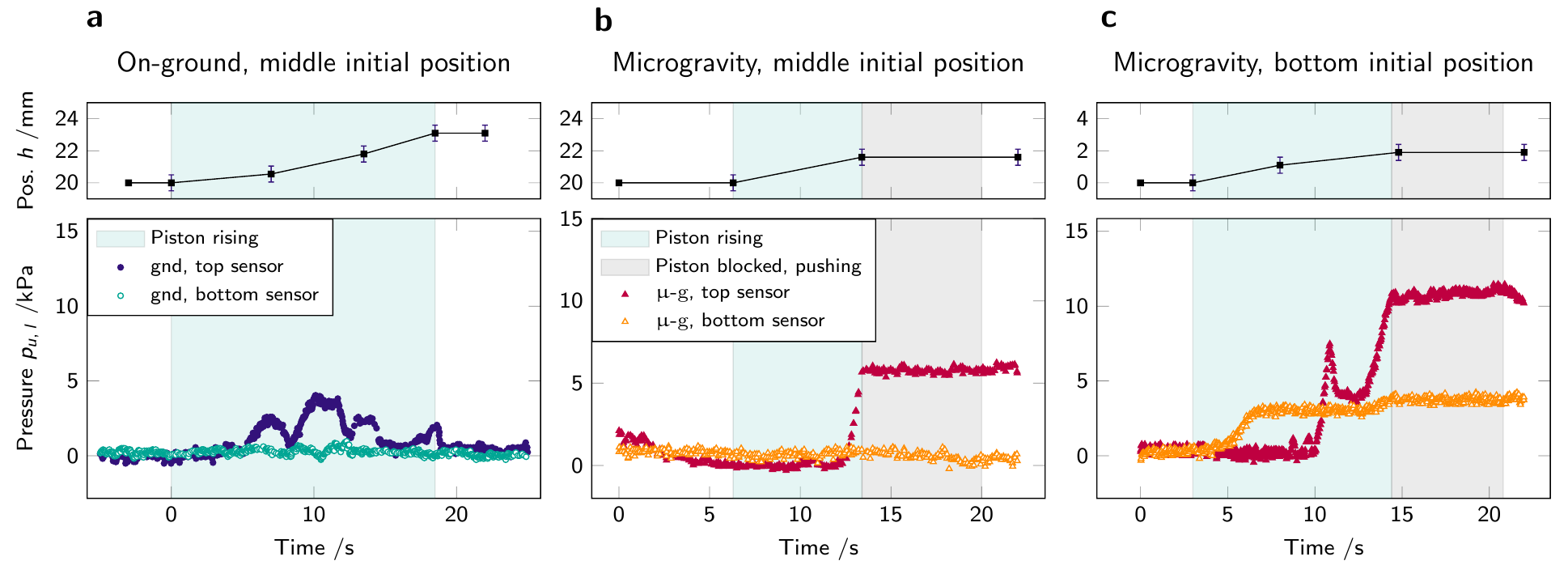}
  \caption{Pressure evolution with piston rise,
(a)~\acrfull{1g} and (b, c)~in \acrfull{mug}. The pressure $p_{u,\,l}$
(bottom graphs) is recorded by sensors in two
positions: on the top part of the cell ($p_u$: filled marks, \gls{1g} in
blue and \gls{mug} in red) and on the bottom part of the cell
($p_l$: open marks, \gls{1g} in green and \gls{mug} in orange).  On (a)
and (b), the piston starts in the middle of the cell (position
$h\approx \SI{20}{\mm}$), while on (c) it starts at the bottom of the
cell ($h = \SI{0}{\mm}$): on (a) and (b) the
pressure sensors are distributed one above the piston
and one below, while on (c) both sensors are above the piston.  
On panel (a),
$t=\SI{0}{\s}$ represents the start of piston rise;
on panels (b) and (c), 
$t=\SI{0}{\s}$ represents the start of the \acrlong{mug} phase.
The pressure evolution exhibits repeated yielding \acrlong{1g}, 
while the packing jammed in \acrlong{mug} remains stable,
even at maximum load.
In the top graphs, the error bars represent the experimental precision in optical measurements.
}
\label{fig:qualitative}
\end{figure*}

\begin{figure*}[h!]
  \centering
\includegraphics[width=\linewidth]{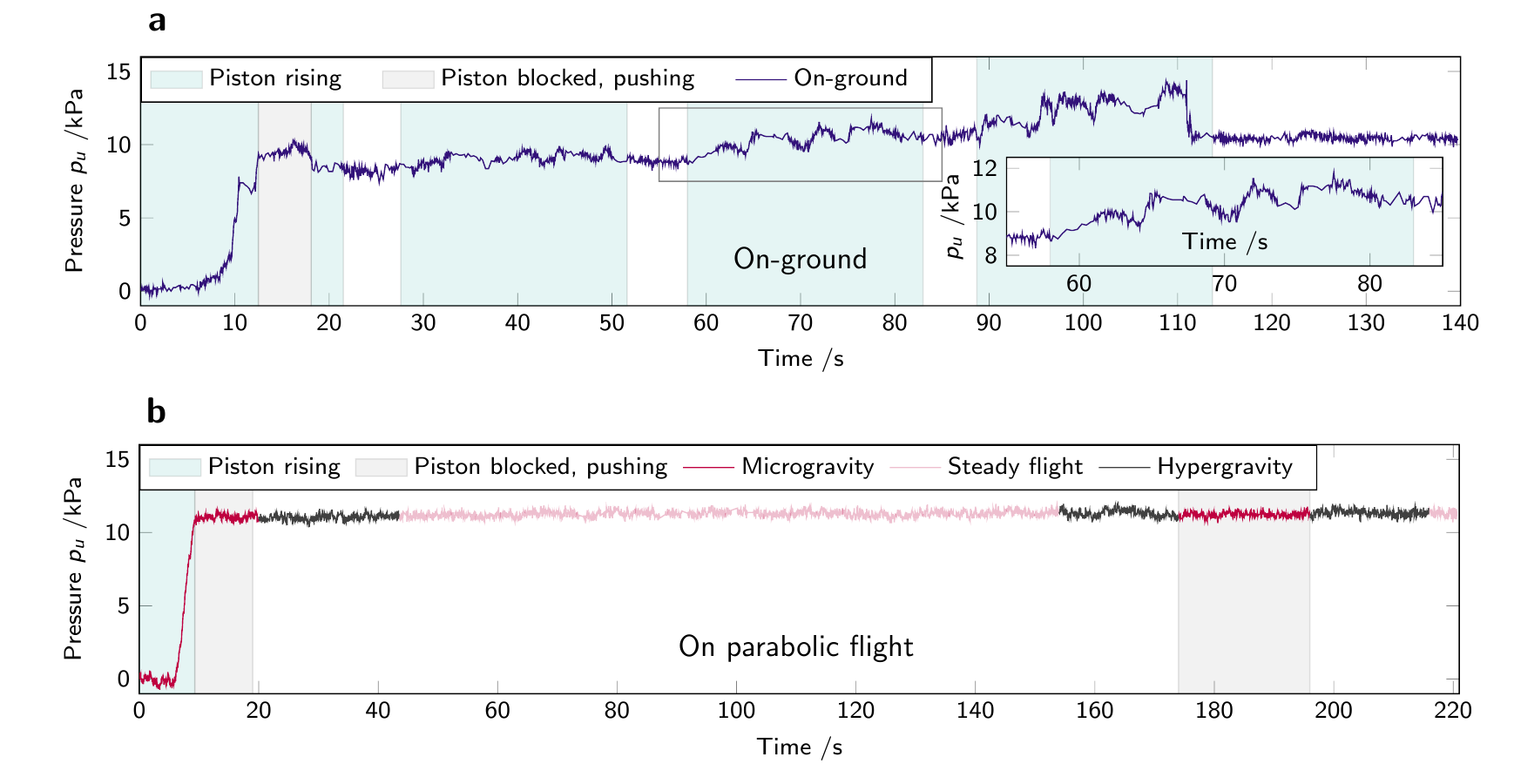}
  \caption{\label{fig:pressure_long} Pressure evolution for a duration of $\approx \SI{200}{\s}$ after the initial
densification step leading to jamming.
Pressure is recorded by the top
sensor inside the powder container, (a)~\acrfull{1g} and (b)~\acrfull{mug}. The inset of
panel~(a) is a close-up of the pressure evolution during the
second step up of the piston, 
where the time and pressure range expanded in the inset are highlighted on the main panel by a grey rectangle.
After an initial steep rise in pressure both \acrlong{1g} and in \acrlong{mug},
corresponding to jamming of the granular packing,
\acrlong{1g} the packing yields under the stress applied by the piston rise.
On the contrary, the packing jammed in \acrlong{mug} remains stable against hypergravity and
against pressure applied by the piston in the next \acrlong{mug} period.
}
\end{figure*}

Above $\varphi_1$, the behavior of the packing depends on the
gravitational environment. \Acrlong{1g}, the powder
undergoes a first large-scale reorganization at
$t^{\text{gnd}} =$~\SI{10.5}{\s} 
(point (iii) in Fig.~\ref{fig:piston_probing_densification}) which temporarily
releases pressure on the sensor, but with almost no powder flowing
under the platform, as seen in
Fig.~\ref{fig:piston_probing_densification}(c) images (iii) and (iv).
The jamming packing
fraction $\varphi^{\text{gnd}}_{\text{J}}$ is reached at
$t^{\text{gnd}}_{\text{J}} =$~\SI{12.5}{\s}.  The position of the piston in height
correlates with the pressure measured on the wall of the container:
the piston is blocked and stops rising when a peak in pressure is reached.  
At this point, the powder enters a creep regime of slow deformation
and eventually yields at $t^{\text{gnd}} =$~\SI{18}{\s}. As a result,
material flows under the platform, releasing the pressure on the
container wall, as seen
on~Fig.~\ref{fig:piston_probing_densification}(c), image (v). By placing the
piston higher in the tube, yielding can be enhanced to the point that
it repeatedly relaxes the pressure completely and allows the piston to
continue to rise (\textit{cf.}~Fig.~\ref{fig:qualitative}(a)). Apart from small
dynamic excursions as material is flowing down, the pressure on the
wall below the piston is unaffected by the developments above the
piston. The creep response may extend over minutes (\textit{cf.}~Fig.~\ref{fig:pressure_long}) with smaller pressure
variations -- indicative of continued internal
rearrangements -- interspersed by yielding events that deposit more
material beneath the piston.

In \acrlong{mug}, the granular behavior is quite different.  
Approximately at the same time as for the ground experiment
($t^{\textrm{$\mu$-g}}=$~\SI{6}{\s}), the pressure rises on the sensor;
however it is not \enquote{jerky} as \acrlong{1g}, but a smooth and relatively
steep increase towards its saturation pressure, indicating an
effectively elastic response of the packing.  The maximum pressure is
reached earlier than on-ground: at $t_J^{\textrm{$\mu$-g}} =$~\SI{9.3}{\s}.
It should also be noted that at this instant 
(point (ii) in
Fig.~\ref{fig:piston_probing_densification}),
despite the platform
having risen by \SI{1.7}{\mm}, almost no powder has \cori{flowed} under the
platform: only the fringe of the piston's platform is slightly covered by less
than~\SI{1}{\mm} of (white) powder.
This is attributed to gravity-jitter in the
horizontal plane, common on parabolic flights~\cite{Carr2018}.

In Fig.~\ref{fig:qualitative},
pressure evolution is presented for one step up of the piston,
\acrlong{1g} (\ref{fig:qualitative}(a))
and in \acrlong{mug} (\ref{fig:qualitative}(b) and \ref{fig:qualitative}(c)).
The pressure $p$
is recorded on the top and bottom part of the container (respectively $p_u$ and $p_l$).
In Fig.~\ref{fig:qualitative}(a) and (b), at
$t = \SI{0}{\s}$ the piston is in the middle of the cell: the top
sensor records the pressure above the piston and the bottom sensor,
that under it.
In particular, for the experiment shown in
Fig.~\ref{fig:qualitative}(b),
the piston, placed at half height of
the cell, is its initial position.
In Fig.~\ref{fig:qualitative}(c),
the initial piston position
is at the bottom of the cell,
hence both sensors are placed above the piston.

While placing the piston higher in the tube does not facilitate
yielding as it does \acrlong{1g},
occasionally, large scale rearrangements can be observed in the
initial buildup of pressure (see respectively Figs.~\ref{fig:qualitative}(b) and \ref{fig:qualitative}(c)).
The latter \cori{Fig.~\ref{fig:qualitative}(c)} also shows that densification proceeds inhomogeneously
throughout the sample.
We observe
that until $t=\SI{10}{\s}$,
the upper sensor does not record any substantial signal change,
while the lower one already undergoes a pressure increase 
of approx.~$\SI{3}{\pascal}$ from
$t=\SI{5}{\s}$,
as the piston starts to rise in the container.
Our conjecture is that
at a point where the lower pressure sensor
already registers substantial stresses, 
no force is yet transmitted to
the upper part of the packing,
as the contact network is not yet formed.

\begin{figure}[h!]
  \centering
\includegraphics[width=\linewidth]{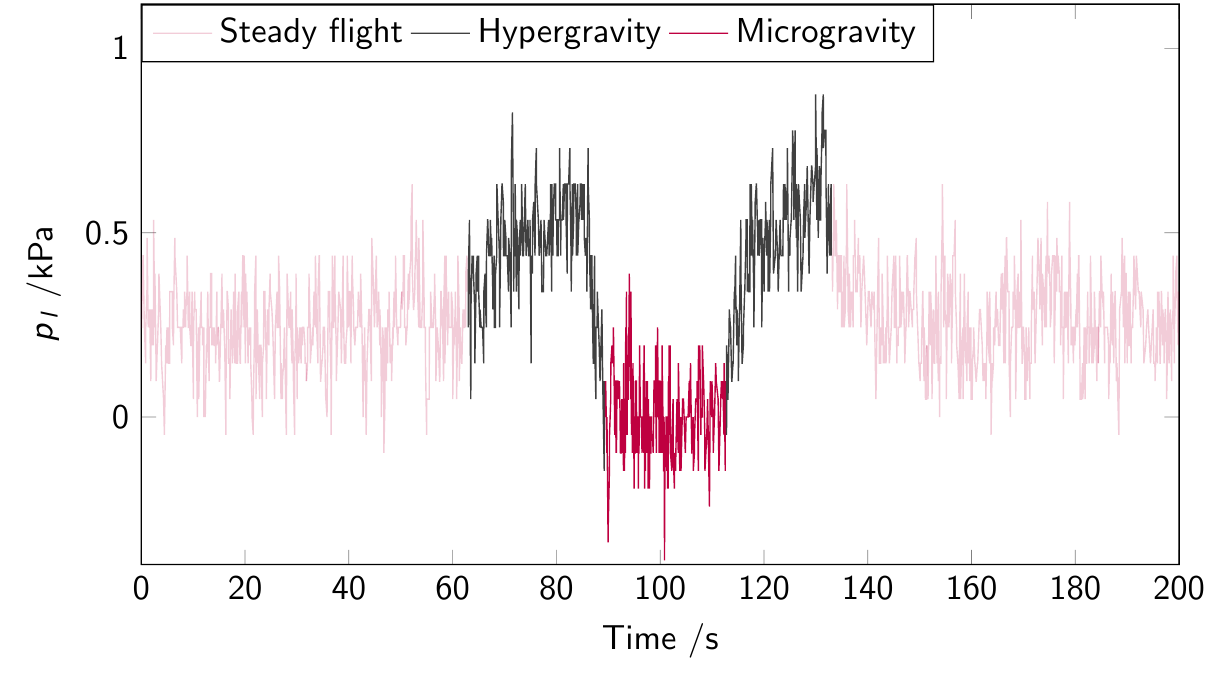}
  \caption{\label{fig:pressure_leftalone}
Pressure $p_l$ on the bottom sensor inside a cell filled
with powder, but where the piston does not move: the pressure
changes are solely due to changes in acceleration
during the parabolic flight maneuvers. Initial time $t=\SI{0}{\s}$ is not a significant instant.
}
\end{figure}

In Fig.~\ref{fig:pressure_long},
the pressure is observed for a longer duration after the first step up of the piston, \acrlong{1g} (Fig.~\ref{fig:pressure_long}(a)) and in \acrlong{mug} (Fig.~\ref{fig:pressure_long}(b)).
\Acrlong{1g}, four repetitions of the piston rise lead to
pressure build-up followed by collapse after $~\SI{5}{\s}$.
The inset is a close-up of the third step up of the piston.
Remarkably, there is no
noticeable creep in \acrlong{mug}. The recorded pressure does not change
through repeated \acrlong{mug} and hypergravity phases. In order to verify
that this is not due to malfunction of the pressure sensors, on a
different flight-day the pressure \cori{was} recorded during mulyiple parabolas
without any motion of the piston
(remaining in its initial position at the bottom of the cell).
The result is presented in Fig.~\ref{fig:pressure_leftalone}. 
If the packing is left in its
initial unjammed state (no compression applied), the pressure
variations due to change in gravitational acceleration is clearly
visible, and corresponds to the temporal structure of the parabolic
flight: hypergravity--low gravity--hypergravity, surrounded by steady flight.
(The reader is referred to Fig.~\ref{fig:crazypressure}
for a longer pressure measurement without piston motion.)

Note that all the experiments reported in
Figures~\ref{fig:piston_probing_densification} to \ref{fig:pressure_long}
involve the same maximum piston load. However, the corresponding
pressure on the side walls depends on the distance between pressure
sensor and piston. In particular, the pressure increases with the
distance to the piston.

To summarize, the granular sample subjected to piston-probing undergoes the
following steps:
\begin{enumerate}
\item Densification of the granular medium: the packing density
  increases on top of the piston, leaving an empty volume under it.
\item The granular medium reaches a critical packing fraction ($\varphi_1$) above the
  piston, dense enough to transmit forces from the rising piston to the sensors on the container walls. Pressure on the side-walls increases and the piston
  slows down.
\item[\refstepcounter{enumi} \number\value{enumi}$^{\,\textrm{$\mu$-g}}$.] In
  \acrlong{mug}, the pressure increases continuously until the piston
  stops at maximum load. 
  We label this point \enquote{jamming point}. 
  Note that this state is stable under
  hypergravity as long as the piston is not lowered.   
\item[\number\value{enumi}$^{\,\text{gnd}}$.] \Acrlong{1g}, the pressure also increases until a maximum load is reached and
  the piston stops. Again, it is this first stop of the piston that we label \enquote{jamming point}. Eventually, the packing yields and flows into the empty space below the piston. 
\end{enumerate}

\subsection*{Identification of regions in the packing}

The piston-probing experiment conducted \acrlong{1g} and in \acrlong{mug}
reveals a gravity-dependent response of the granular material on 
two main aspects:
\begin{itemize}[label={-}]
    \item Under compression, jamming occurs in \acrlong{mug} at lower packing fraction than \acrlong{1g}.
    \item The packing eventually yields and flows \acrlong{1g}, due to the secondary (gravitational) force field; in its absence, i.e.~in \acrlong{mug}, the granular material arranges into a highly stable jammed state.
\end{itemize}

\begin{figure*}[h]
  \centering
\includegraphics[width=0.93\textwidth,trim=0mm 1mm 0mm 0mm,  clip]{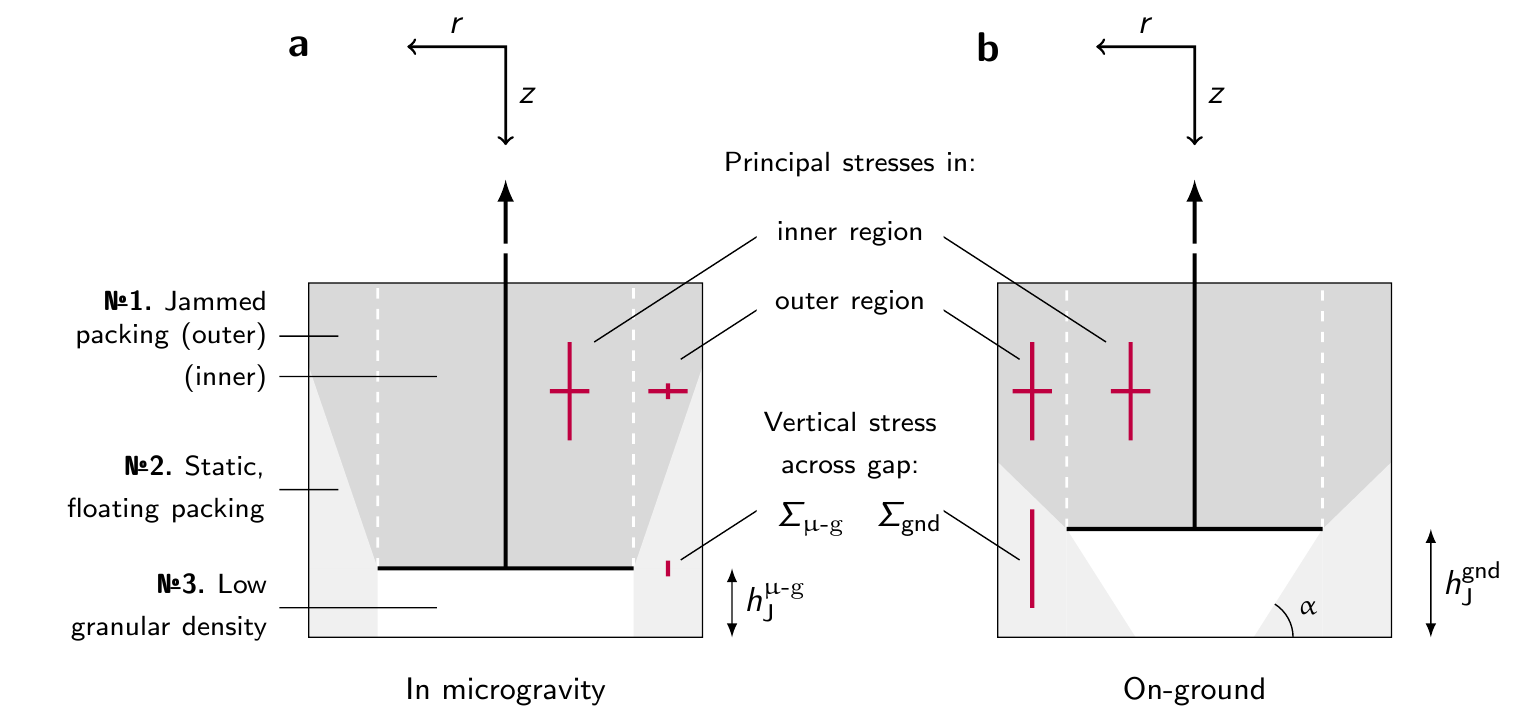}
\caption{Schematic representation of the regions appearing in the container as the piston rises,
(a)~in \acrfull{mug} and (b)~\acrfull{1g}. \Acrlong{1g}, the
    powder forms a slope at its angle of repose $\alpha$ with the
    horizontal. Stress magnitudes in arbitrary units are depicted by
red lines. The coordinate system used in
Sec.~\ref{sec:stress-distr-region-1} is shown above each schematic. 
The height $h_J^{\textrm{$\mu$-g}, \, \text{gnd}}$ denotes the height at which the piston is stopped by the jammed material above it, respectively under \acrlong{mug} and \acrlong{1g} conditions.     
    The difference in force chain configuration between \acrlong{mug} and ground leads to drastically different vertical stresses in the outer region. The load $\Sigma$ on the material below the piston is, as a consequence, much lower in \acrlong{mug} compared to ground.
See also Supplementary Material, Supplementary Figure~1.}
  \label{fig:regions_pistonbox}
\end{figure*}

To understand the granular
physics behind the observed behavior and explain the gravity-dependence, 
we need to have a closer look at the granular material
in the apparatus. To this end, it is helpful to conceptually
distinguish three different regions inside the tube,
represented in Fig.~\ref{fig:regions_pistonbox}.
Region \textnumero1 is the
material above the piston, compacted as the piston rises; it is the region where the granular material eventually jams. 
Region \textnumero3 is the
region below the piston, left empty as the piston rises.
In between, region \textnumero2 is the region around the piston,
which can flow into the empty region \textnumero3.

The task ahead can therefore be structured into: (1) understanding the
compaction phase of region \textnumero1 and the resulting granular fabric; (2)
understanding the stress distribution in region \textnumero1 that
determines the force on region \textnumero2; and, finally, (3),
understanding the yield criteria of region \textnumero2 in the
presence of an empty region \textnumero3.

\subsection*{Compaction of region \textnumero1}
\label{sec:comp-region-1}

While characterizing the point of transition to jamming is still a
controversial topic~\cite{Makse2004, OHern2002, Jerkins2008, Chaudhuri2010, Behringer2018}, the mesoscopic scale phenomenon at its origin is
generally accepted: compression-induced jamming occurs if strong force chains appear along
the major compression stress direction, in response to the load
itself~\cite{Farr1997}.  The presence or absence of a secondary
(gravitational) force field modifies the force distribution in the
different zones of the packing~\cite{Murdoch2013}.

At $\varphi_0$, prepared under ground conditions, the packing
assumes a configuration that is rigid against the unidirectional
gravitational body force, 
as long as no other load is applied. This structure, however, is no longer rigid
against the unidirectional boundary force applied by the piston,
and the
packing can initially be compacted with negligible force.

Due to frictional interactions, vertical forces are
partially supported by the vertical outer walls of the tube. Force
chains, consequently, arch in the direction of the net force. When the
latter changes from the downward orientation due to gravity, to the
upward orientation imposed by the piston, 
we can infer that
the force chains
change curvature. Where the forces balance, the force chains
buckle. For cohesionless particles, this would in itself allow the
packing to further collapse under the influence of gravity as the
unloaded contacts between particles are no longer constrained by
friction and/or cohesive interactions.

At this point it is instructive to estimate the maximal stress imposed
by gravity \acrlong{1g}.
From the analysis proposed by \citeauthor{Janssen1895}~\cite{Janssen1895}
(\textit{cf.}~Sec.~\ref{sec:stress-distr-region-1}),
we find a screening length
$\ell\approx\SI{26}{\milli\metre}$, which limits the gravitational
stress even at maximum depth $H$ to $\sigma_g \approx \SI{125}{\pascal}$
(\textit{cf.}~Eq.~\ref{eq:6}). As seen below in Sec.~\ref{sec:stab-region-2},
the cohesive energy density is at least on the order of
\SI{10}{\kilo\joule\per\cubic\metre},
(equivalent to \SI{10}{\kilo\pascal}, as
stress and energy density have the same units).
Typical forces during buckling
of the force chains will therefore be unable to break most
contacts. The new 
network of force chains will initially form by mostly
establishing additional contacts between particles, which necessarily
implies a densification of the packing.

As a result of the densification, 
$\varphi_1$ is eventually reached. 
We can infer the following mechanism on the microscopic scale:
the granular fabric supports load-bearing force chains again, 
as the chains of contacts 
above of the piston have percolated into a network transmitting the load imposed by the piston to the container's walls.
However, this network is not stable against the piston-imposed load:
the particles can still reorganize into denser configurations.

As the piston does not cover the whole cross-section of the tube, 
we may conceptually distinguish
between an \emph{inner} region \textnumero1 and an \emph{outer} region
\textnumero1. The inner region directly transmits the forces of the
piston, while the outer region is {indirectly} affected through the
horizontal redistribution of stresses in the packing. Under gravity,
as soon as the stress due to the piston rise exceeds the gravitational stress,
forces are directed upwards while the outer region may still be forced
down by gravity. This create a shear force between inner and outer
that intermittently severs force chains. The resulting local force
balance may lead to large scale rearrangements that are at the origin
of the pressure drops observed \acrlong{1g} between $\varphi_1$ and
$\varphi_{\text{J}}$. Note that the stresses imposed by the piston considerably
exceed the gravitational stresses in the inner region for most of the
compression phase such that gravity is negligible there.

At $\varphi_{\text{J}}$, the packing has found a configuration that is stable
against the maximal piston force with respect to local rearrangements
of particles.
On longer time scales however, 
for the experiments conducted \acrlong{1g},
the evolving pressure
signal indicates an ongoing evolution of the force chain network under
constant load (see Fig.~\ref{fig:pressure_long}(a)). 
Note that the pressure on the side walls is increasing
over time, which implies that the deflection of forces away from the
vertical also increases and the vertical normal stress decreases in
the upper parts of the piston. This aging or creep phenomenon can be
rationalized in terms of \gls{sgr}~\cite{Sollich1997,Fielding2000}
and its adaption to jammed packings \cite{Bi2009}. On a continuum
level, force chains entail a spatially heterogeneous stress
distribution. Global stress relaxations may therefore occur by chains
of local relaxations that proceed at a finite speed. \cori{In the
framework of Edward's ensemble of granular packing, the \emph{angoricity}
is the intensive thermodynamic parameter conjugate to the stress; it
quantifies the sensitivity of the stress heterogeneities to the
imposed load \cite{Blumenfeld2009}.} Within
\gls{sgr}, jamming occurs if 
the angoricity $\beta$
falls below a critical value $\beta_J$. In the
jammed state, $\beta<\beta_J$, \gls{sgr} predicts aging
\cite{Fielding2000}, compatible with our measurements.

For the \acrlong{mug} experiments, 
upon entering the \acrlong{mug} phase of
each parabola, the cohesive forces among particles become
predominant~\cite{Love2014} and create a packing where
once particles come into contact with their closest neigbours,
those contacts are likely to remain.
Upon compression and up to $\varphi_1$, the packing
collapses -- like \acrlong{1g}, mostly by local buckling and by
preserving existing contacts -- into a load bearing
configuration. However, no strong shear forces appear as the piston is
the single source of stress \coriii{on the mesoscopic scale}. Note that friction with the piston's
wall is generically smaller than internal friction of the granular
packing~\cite{Nedderman1992}. At the slow speed of the piston, the
wall friction therefore cannot generate shear forces between the inner
and outer region. Remarkably, under these conditions, the topology of
the force chain network seems to be independent of the magnitude of
the imposed stress. As a result, the measured pressure $p_{u, \, l}$ becomes
proportional to the strain imposed by the piston and the packing
behaves as an elastic solid with a well defined elastic modulus. As a
rough estimate from Fig.~\ref{fig:piston_probing_densification}, let's
assume a strain of 1\% during the buildup of pressure from
\SI{0}{\kilo\pascal} to \SI{10}{\kilo\pascal}. This yields an elastic
modulus on the order of \SI{1}{\mega\pascal}, compatible with
theoretical predictions (see Methods Section).

On the time scale of the experiment, no creep can be observed in
\acrlong{mug}. In terms of the angoricity, this implies a vanishing
sensitivity to the magnitude of the load, $\beta\to0$. This is in
line with the elastic regime observed before for $\varphi <
\varphi_J$. At this stage we can only deduce that a single boundary
force can lead to $\beta=0$ while the competition between said
boundary force and the gravitational body force leads to a finite
$\beta > 0$. We invite further studies that investigate this more
systematically.

A \gls{2d} hands-on version of a section of this experiment using photo-elastic
discs allowed us to visualize the force chains distribution within the
granular medium; it qualitatively confirms the appearance of the three
regions sketched in Fig.~\ref{fig:regions_pistonbox}. Results are
shown in Supplementary Material (Supplementary Figure~1).

\subsection*{Stress distribution in region \textnumero1}
\label{sec:stress-distr-region-1}

In order to understand the load the piston is indirectly enforcing on
region \textnumero2, 
one has to understand the stress distribution in
region \textnumero1. 
Of particular interest is
the normal
stress $\Sigma$ imposed by region \textnumero1 to region \textnumero2
through the gap around the piston.

The observation that the vertical free surface of the granular
material below the piston is unstable at heights of the order of
millimeters if prepared \acrlong{1g}, 
yet remains stable even under
hypergravity conditions if prepared in \acrlong{mug},
is perplexing, at first. It is testament to the nontrivial stress
distribution in a granular sample. In the following, 
we will \coriii{propose a
continuum argument to rationalize this observation.}

Following Janssen \cite{Janssen1895,Nedderman1992} (see
Sec.~\ref{sec:method}), one may, for the sake of the argument, assume
that the vertical, $\Sigma_{zz}$, and horizontal, $\Sigma_{rr}$,
normal stresses are the principal stresses. Unlike in a molecular
fluid, the two normal stresses are not identical, even in the absence
of gravity, and are not homogeneous throughout the material. Again
following Janssen, we restrict our discussion to a two dimensional,
vertical cross section of the experiment (which is in principle trivial to extend to \acrshort{3d}, 
given the axisymmetry of our experimental setup).

At finite gravity but with an immobile piston, the experiment is
equivalent to the silo Janssen had in mind. We consider a silo with a circular cross-section of diameter $D$; in summary, the major
principal stress is $\Sigma_{zz}$ and the minor principal stress,
$\Sigma_{rr} = k\Sigma_{zz}$, is reduced by Janssen's constant $k < 1$.
Friction with the wall will screen the weight on a length scale
$D/4\mu k$.
Here, we focus on the radial rather than the vertical
structure of the problem, and just keep in mind that from one end to
the cylinder to the other, stresses will be renormalized by the
screening effect.

Let's begin by recalling that the normal stress due to gravity on the
gap around the piston,
$\Sigma\equiv\Sigma_{zz}(z=H) \leq \sigma_g \approx \SI{125}{Pa}$. 
Now we
imagine a significant normal stress $\sigma\gg \sigma_g$ being applied
by the piston, such that we can neglect gravitational stresses in the
following. In addition, consider only the material directly above the
piston, for $r < d/2$. The Janssen effect will screen the stresses on a
length scale $\ell = d/4\mu k$, leading to a depth averaged vertical
normal stress $\overline\Sigma^<_{zz} = C_1\sigma$ and related to
this, an average horizontal normal stress
$\overline\Sigma^<_{rr} = k\overline\Sigma^<_{zz}$. Here
$C_1 = C_1(H/\ell) < 1$ is a dimensionless constant encoding the
geometry of the setup. From Fig.~\ref{fig:qualitative}(c) we infer
that the pressure between at the two heights differs by about a factor
$2.5$ in the fully jammed state. Given the distance between the
sensors (see Fig.~\ref{fig:piston_probing_setup}), we obtain an estimate
for the screening length, $\ell\approx\SI{26}{\milli\metre}$.

If the packing is prepared \acrlong{1g}, the major principal stress
will also be vertical in the outer region, $D/2 > r > d/2$, but the
geometry of the outer region will induce its own screening length
$L\approx D$. If we assume the applied stress to dominate over gravity,
$\sigma \gg \sigma_g$, then in the outer
region\footnote{$C_2(L) = C_2(H/\ell, H/L, \ell/L) < 1$ is another
  dimensionless geometrical factor.}
$\overline\Sigma^>_{zz} = C_2(L)\overline\Sigma^<_{rr}/k$, while
$\overline\Sigma^>_{rr}(r) = \overline\Sigma^<_{rr}d/2r$. In
particular, the pressure recorded on the side wall
$p \approx \overline\Sigma^>_{rr}(D/2) = (C_1kd/D)\sigma$ and the stress
on the gap
$\Sigma_{\text{gnd}} \approx \overline\Sigma^>_{zz} = C_2(L)C_1\sigma$.  From
this we conclude $\Sigma_{\text{gnd}} \approx p/k$. Up to renormalizations, the
normal stress on the gap is essentially given by the stress applied
through the piston. The pressure recorded on the side wall, on the
other hand, is proportional to $k\sigma$.

However, if prior to applying a stress, the sample is prepared in
\acrlong{mug} conditions, the compaction in the outer region is
effected by the horizontal normal stress, $\Sigma^<_{rr}$, exerted by
the inner region. Therefore the major principal stress is now
horizontal instead of vertical in the outer region and only a fraction
$k$ is transmitted to the vertical direction,
$\Sigma^>_{zz} = k\Sigma^>_{rr}$. This will shorten the screening
length in the outer region to $L' = k^2L$. For the average vertical
normal stress we now have $\overline\Sigma^>_{zz} =
kC_2(L')\overline\Sigma^<_{rr}$, i.e., $\Sigma_{\textrm{$\mu$-g}} \approx
\overline\Sigma^>_{zz} = k^2C_2(L')C_1\sigma$, while the pressure on
the side wall remains essentially unchanged. From this we conclude
$\Sigma_{\textrm{$\mu$-g}} \approx kp$ and most importantly $\Sigma_{\textrm{$\mu$-g}}/\Sigma_{\text{gnd}} \approx k^2$ independent of the applied load.

Notably, the load on the packing below the piston is significantly
reduced under \acrlong{mug} conditions. The additional weight induced
stress in a subsequent hypergravity phase is heavily screened by the
short length scale $L' \approx k^2D \approx \SI{9}{\milli\metre}$, to
roughly $\gamma L' \lesssim \SI{50}{\pascal}$. This is well within the
stress fluctuations due to g-jitter. If the packing was stable to the
latter, it will not yield in hypergravity either.  As a rough
estimate, we infer from the maximum $p\approx\SI{11}{\kilo\pascal}$,
\begin{subequations}
  \begin{align}
    \label{eq:7}
    \Sigma_{\text{gnd}}&\approx \sigma \approx \frac{p}{k\left(1 -
                 e^{-H'/\ell}\right)} &\approx \SI{40}{\kilo\pascal},\\
    \intertext{and}
    \Sigma_{\textrm{$\mu$-g}}&\approx k^2\sigma \approx \frac{kp}{\left(1 -
                 e^{-H'/\ell}\right)}&\approx \SI{6.5}{\kilo\pascal},
  \end{align}
\end{subequations}
where $H'=\SI{30}{\milli\metre}$ is the approximate distance between
the upper surface of the piston and the pressure sensor at maximum
load.

Note that Mohr-Coulomb theory has nothing to say on the creep
behavior. From the slowly growing horizontal stress at constant
vertical load, we may infer that Janssen's constant is also slowly
growing in time and the packing evolves towards a slightly more
isotropic stress distribution.

\subsection*{Stability of region \textnumero2}
\label{sec:stab-region-2}

First of all note that a vertical free surface of a granular packing
can only be stable due to cohesion, $c > 0$. Effective values for the
cohesive stress (or energy density) are hard to come by and do not
only depend on the particles but also on the preparation. However,
typical values for ordinary materials are on the order of several
kilopascal. Under gravity, the weight stress $\gamma h$ limits the
maximum stable height of a vertical surface to
$h_{\mathrm{max}} = 4c/\gamma(\sqrt{1 + \mu^2} - \mu)$
\cite{Nedderman1992}. With
$\gamma_{\text{gnd}} \approx \SI{6}{\kilo\pascal\per\metre}$, one obtains stable
heights on the order of meters. It is obvious that we cannot expect
collapse of the free surfaces of a couple millimeters in height that
occur in the experiment even under hypergravity. In addition, we have
seen above that the load imposed by the piston on the gap, even under
the unfavorable \acrlong{mug} conditions, by far exceeds the gravitational
load.

Mohr-Coulomb theory also provides the maximum load that can be
sustained by a vertical wall,
$\Sigma_{\mathrm{max}} = 2c/(\sqrt{1 + \mu^2} - \mu)$
\cite{Nedderman1992}. Given that the experiment yields \acrlong{1g} but not in \acrlong{mug}, we can deduce a cohesion strength
on the order of a few kilopascal in line with expectations. Once
region \textnumero2 yields under ground conditions, the material will
settle in a heap with angle of repose $\alpha$ determined by the
material properties. This limits the amount of material that can be
deposited below the piston, as alluded to in the introduction.

\subsection*{Shift of the density at jamming}
\label{sec:shift-jamming-point}

Concomitantly to the maximum height reached by the piston under the
different experimental environments, the critical packing fraction at
which jamming occurs under compression is found to be lower in
\acrlong{mug} \corii{(on average $\varphi^{\textrm{$\mu$-g}}_{\text{J}} = 0.567 \pm 0.014$) than \acrlong{1g}
(on average $\varphi^{\text{gnd}}_{\text{J}} = 0.579 \pm 0.014$)} (see Fig.~\ref{fig:critical_height}).

Variability in the jamming packing fraction $\varphi_{\text{J}}$ 
has notably been studied
for increased frictional interactions between particles,
which was linked
to a reduction in the packing fraction at jamming~\cite{Wyart2014, Mari2014}: 
\cori{if the particles' contact number is}
high enough, the jamming
transition happens at lower packing fraction for frictional particles
than for frictionless particles.
The proposed mechanism is the following:
increased frictional interactions
restrict the reorganization possibilities of the packing, resulting in
a higher stability of each configuration.
In other words, the
reduction of collective degrees of freedom available for the granular
fabric to reorganize into a denser packing under the applied stress
results in a jamming transition happening at lower packing fraction.
Similarly, it is our hypothesis that the lower packing fraction at
jamming observed in \acrlong{mug} results from a decrease in collective
degrees of freedom and manifested in a reduced spatial heterogeneity
of the force network.

The absence of the force field due to Earth gravity
leads to another related phenomenon: 
the cohesive interparticle forces, 
insignificant when compared to the particles' weight, 
become predominant in \acrlong{mug}.
Stronger cohesive interactions 
also reduce the freedom of motion of each individual particles,
hence of the granular packing to reorganize into a denser configuration as a whole.

This observation confirms results from one of the first rheology
experiments performed in \gls{mug}, on board the Space Shuttle
\cite{Sture.1998, Alshibli2003}. 
A startup curve was measured for a
dense ($\varphi=0.65$), polydisperse powder, comparing \gls{1g} and
\gls{mug} conditions.  The response in \gls{mug} was jagged, and samples in \gls{mug}
exhibited a much higher peak friction strength.

\section*{Discussion}\label{sec:conclusion}

A granular piston-probing experiment was conducted \acrfull{1g} and in
\acrfull{mug} (using parabolic flights as a low-gravity platform) to
probe the effect of the gravitational acceleration on the compression of a
granular medium. We find that the collective behavior of a granular
material reveals strong dependence on the gravitational environment.

As the piston rising densifies the granular material above it
(region \textnumero1),
an equivalent empty volume is created below the piston
(region \textnumero3).
Once the material in region \textnumero1 reaches a 
packing sufficiently dense to transmit the load imposed by the piston, 
a rise of pressure recorded on the container's walls indicates
the formation of a percolated network,
and the piston upward motion slows down.
Eventually, a jammed state is reached above the piston (region
\textnumero1): the load imposed by the piston is transmitted to the
container's walls and eventually the piston stops. \Acrlong{1g}, the fabric of
the jammed packing allows for sufficient downward forces to overcome
cohesion in creep flow. The formation of an annular heap with a finite
angle of repose in region \textnumero3 eventually limits the powder
rearrangement and jams the piston. In \acrlong{mug}, the fabric is
altered in such a way, that the downward forces are insufficient to
overcome cohesion. The piston jams even though region \textnumero3
still provides free space.

In our experiments conducted in \acrlong{mug}, this jammed state is
reached while the piston is at lower height,
$h_{\text{J}}^{\textrm{$\mu$-g}} < h_{\text{J}}^{\text{gnd}}$, than for their repetitions
\acrlong{1g}. We conclude that the packing fraction at jamming is
lower in \acrlong{mug} than \acrlong{1g}. 
We support this finding by two \cori{physical
explanations. Firstly,} as the interparticular force balance changes in absence of
gravity, cohesive interactions become predominant. 
\cori{Besides, \acrlong{1g}, 
particles minimize their potential energy 
by going to the lower position possible:
Earth gravity creates a
secondary force field to stimulate reorganization in a
different direction to the major compression stress. 
Under \acrlong{mug}, such reorganization stimulation does not take place.
As a result, in
\acrlong{mug}, powder flow deteriorates; 
the particles' configuration is more stable
or less likely to reorganize:}
the packing fraction at jamming is lower in
\acrlong{mug} than \acrlong{1g}.

In all experimental results presented,
the piston height $h$ is the parameter measured directly:
the \emph{global} packing fraction $\varphi$ is 
calculated assuming density homogeneity in the container.
Local packing fractions are not experimentally accessible in the current setup. 
We also report
a lack of control over the initial state of the granular packing,
as the flight time preliminary to the first parabola and repeated hypergravity phases
might contribute to create an initial state specific to the \acrlong{mug} experiment,
that is not reproduced in their \acrlong{1g} counterpart.
To minimize 
the influence of repeated hypergravity phases on the granular packing, 
only the first parabolas of each flight are used to calculate $\varphi_{\text{J}}$.
Besides, the granular samples
generally undergo
a slight positive acceleration along the $z$-axis at the initial instant of each parabola,
which is expected to counteract the compaction caused by hypergravity.

The authors hope that those results will spur interest and further
investigation on the influence of gravity on the packing fraction at
jamming in granular media.
If confirmed,
the deterioration of powder flow 
and lower packing fraction at jamming
under reduced gravity
could have dramatic consequences for 
powder handling on 
sand-covered celestial bodies.

\section*{Methods}\label{sec:method}
\subsection*{Granular material}
\begin{figure}[h!]
\centering
\includegraphics[width=\linewidth]{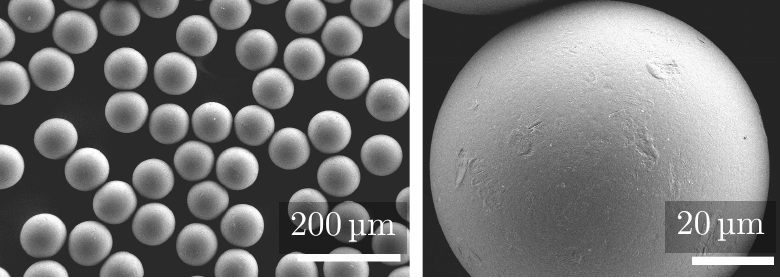}
\caption{\label{fig:micrographs} 
Scanning electron microscopy of the granular material used. The micrographs presented are taken after completion of the cleaning process.
}
\end{figure}

The granular material used for all experimental results presented is a
spherical, monodispersed \gls{ps} powder of mass density
$\rho_b=\SI{1050}{\kilogram\per\cubic\metre}$ and diameter
$a=\SI{80}{\micro\meter}$ produced by the company \textit{Microbeads}
under the name \textit{Dynoseeds}~\cite{DynoseedsMSDS}.
The powder was washed to ensure a clean, smooth surface, and sieved to
avoid outliers
(wet-sieving under ultrasound at frequency of 
\SI{20}{\kilo\hertz} for a duration of 
8 hours).
The surface state and sphericity of the granular material was verified using scanning
electron microscopy.
Micrographs before and after the cleaning process were taken, showing 
the limited roughness and clean surface of the particles after cleaning.
Fig.~\ref{fig:micrographs} shows the cleaned particles, as used in the experiments discussed.

\subsection*{Bulk modulus of the packing: limits to stresses and speeds}
For elastic particles in a given jammed packing, a further reduction
of volume, $\Delta V$, can be achieved by forcing the particles to
deform. Assuming an isotropic compression leads to small typical
deformations $\langle\xi\rangle/a$ relative to the particle radius $a$,
we can estimate the relative volume change as
$-\Delta V/V \approx 3\langle\xi\rangle/a$. Such a volume change is
related to the pressure $\Delta p = -K\Delta V/V$ by the bulk modulus
$K$. \citeauthor{Shaebani2012}~\cite{Shaebani2012} have shown that
\begin{equation}
  \label{eq:1}
  K = C\frac{\varphi z}{6\pi a}\times\frac{\partial F}{\partial
    \xi}(\langle\xi\rangle),
\end{equation}
where $z$ is the average coordination number, $F(\xi)$ is the
inter-particle force, and $C$ is of order one for almost monodisperse
packings. Here we assume a nonlinear, Hertzian contact force
\cite{Hertz1882}
\begin{equation}
  \label{eq:2}
  F(\xi) = \frac{\sqrt2}{3}\times\frac{E_p}{1 - \nu^2}a^{1/2}\xi^{3/2}
\end{equation}
that is parametrized in terms of $E_p$ and $\nu$, Young's modulus and
Poisson's ratio of the particles' material, respectively. Using
Eq.~(\ref{eq:2}) in Eq.~(\ref{eq:1}), and noting that
$\varphi z/(1 - \nu^2) \approx 4$ around jamming where $\varphi \approx 0.6$,
$z \approx 6$, and for typical Poisson's ratios $\nu \approx 0.35$, we find
\begin{equation}
  \label{eq:4}
  K \approx \frac{E_p}{3\pi}\sqrt{2\langle\xi\rangle/a} 
  \approx \frac{K_p}{\varphi}\sqrt{\langle\xi\rangle/128a},
\end{equation}
where $K_p$ is the bulk modulus of a particle, and
\begin{equation}
  \label{eq:3}
  \frac{\Delta p}{E} 
  \approx \frac{\sqrt2}{\pi}\left(\langle\xi\rangle/a\right)^{3/2}.
\end{equation}
If we want to limit the deformations of the \gls{ps} particles
($E(\mathrm{PS}) \approx\SI{3.3}{\giga\pascal}$~\cite{Wypych2016}),
to $\langle\xi\rangle/a\lesssim 10^{-3}$, we need to limit the pressure
to $\Delta p \lesssim \SI{45}{\kilo\pascal}$.

With the universal relation $G = 3K/5$ \cite{Shaebani2012} for the shear modulus in a packing, the s-wave speed
$c_s = \sqrt{G/\rho}$, can be related to the speed of sound in a
particle, $c_p= \sqrt{3K_p(1 - \nu)/\rho_p(1 + \nu)}$ \cite{mavko2009}
($c_p(\mathrm{PS})\approx\SI{2350}{\metre\per\second}$ \cite{Wypych2016}) by Eq.~(\ref{eq:4}) as:
\begin{equation}
    \label{eq:8}
    c_s \approx c_p\sqrt{\frac{1 + \nu}{5(1 - \nu)}}(\langle\xi\rangle/128a)^{1/4}
    \approx \SI{80}{\metre\per\second},
  \end{equation}
much larger than the piston speed. Note that these are order of magnitude
estimates only and many quantitative refinements would be possible.

\subsection*{Mohr-Coulomb material}
Following \citeauthor{Nedderman1992}~\cite{Nedderman1992}, a two
dimensional Mohr-Coulomb material is characterized by three
parameters: the macroscopic friction $\mu$, the cohesion $c$, and the
weight density $\gamma = \rho^*g$ where $\rho^* = \rho_b\varphi$ is
the mass density of the material at a given packing fraction
$\varphi$ (where $\rho_b$ is the density of the particles' material). The Mohr-Coulomb failure criterion limits the shear
stress $\tau$ for a stable material depending on the normal stress
$\Sigma$, $|\tau|\leq \mu\Sigma + c$. One consequence of this yield
criterion is that normal stresses are not isotropic but that there is
a finite ratio $k<1$ between the minor and the major principal normal
stress, the Janssen constant.

Assuming that the major and minor principal stress axes are aligned
with the vertical and horizontal orientation of the container, one
obtains the vertical force balance
\cite{Janssen1895,Nedderman1992}
\begin{equation}
  \label{eq:5}
  \frac{d\Sigma_{zz}}{dz} \pm \frac{\Sigma_{zz}}{\lambda} = \gamma
\end{equation}
and the boundary condition $\Sigma_{zz}(h) = \sigma$ for an
external stress $\sigma$ applied at height $z=h$. The sign between the
first and the second term depend on the direction of the force. The
length scale $\lambda$, associated with the screening of forces
described by Eq.~(\ref{eq:5}), depends on the geometry of the
problem. For a major principal stress in the vertical orientation (the
classical Janssen analysis), $\lambda\propto1/k$, whereas for a
horizontal major stress, $\lambda\propto k$. 
For a container of width $D$ and no external stress, Janssen
\cite{Janssen1895} obtained for the horizontal stress
\begin{equation}
  \label{eq:6}
  \Sigma_{rr}(z) = \frac{\gamma D}{4\mu}\left(1 - e^{-4\mu kz/D}\right)
\end{equation}
and for the vertical stress $\Sigma_{zz}(z) = \Sigma_{rr}(z)/k$.

\subsection*{Data processing}
\begin{figure}[h]
  \includegraphics[width=\linewidth]{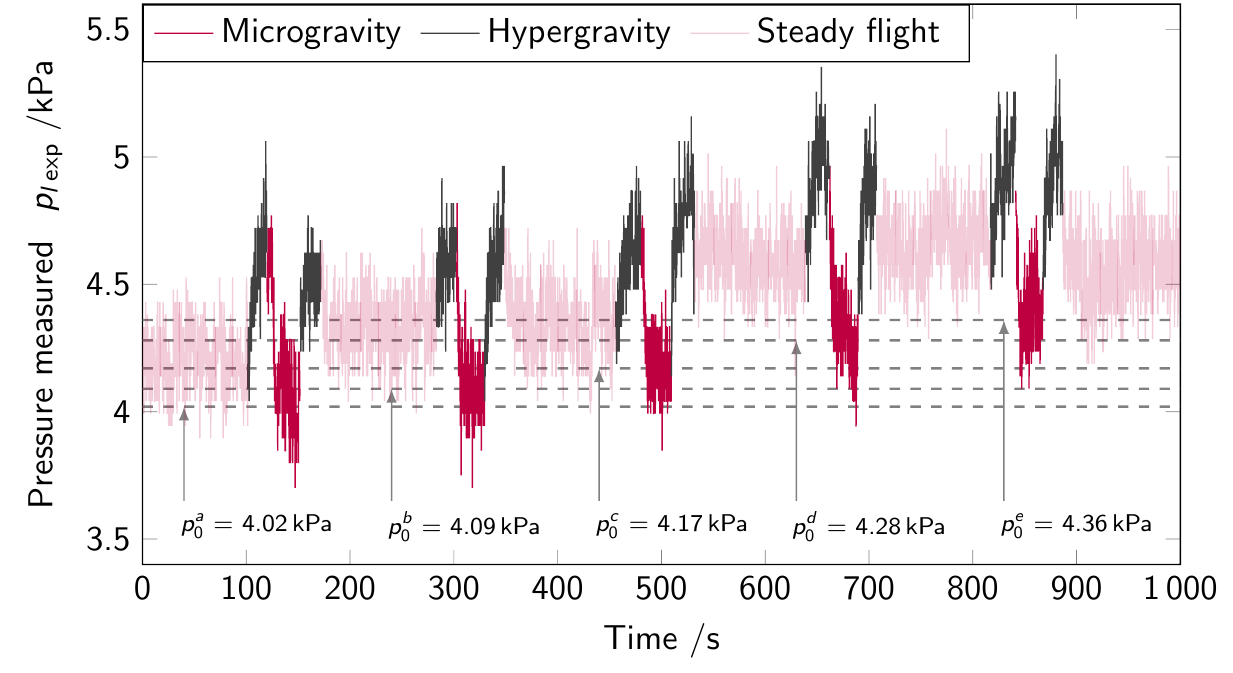}
  \caption{\label{fig:crazypressure} Pressure $p_{\text{exp}}$ measured on
    the bottom sensor inside an experimental cell filled with powder
    but with no motion of the piston. Pressure changes are solely due
    to changes in acceleration during the parabolic
    flight maneuvers. The offset pressures found from the \acrlong{mug}
    phases are shown as dashed lines, labeled
    $p_0^{a, \, b, \, c, \, d, \, e}$. The increase in offset pressure
    throughout the experiments showed here is clearly visible,
    amounting to $+\SI{0.34}{\kilo\pascal}$ in \SI{1000}{\s}.}
\end{figure}

The pressure sensors placed inside the container were observed to
offset
due to the constant pressure applied on the sensors by the granular
packing, once the material is placed inside the container.
This offset was found to be around \SIrange{4}{5}{\kilo\pascal} on all
experiments (both \acrlong{1g} and in \acrlong{mug}).  However, this
offset pressure is not constant, but increases with the experiment's
duration, rendering exact quantitative pressure measurement
challenging.  This shift can be observed in
Fig.~\ref{fig:crazypressure}, where the raw pressure measured
$p_{\text{exp}}$ is shown: throughout the five parabolas described by the
Zero-g plane in the course of \SI{1000}{\s}, the pressure increases by approximately $+\SI{0.34}{\kilo\pascal}$.

In order to retrieve the pressure evolution $p$, from the pressure
measured $p_{\text{exp}} = p + p_{0}$, the offset pressure $p_{0}$ was
calculated per experiment (per parabola in the case of \acrlong{mug}
experiments).

\begin{table*}[h!]
  \centering
  \caption{Numerical values of the parameters describing our system: a \acrlong{ps} powder in \cori{a} \acrlong{pc} container. 
\cori{A plethora of ways are available to estimate the Janssen coefficient $k \in [0, \,1]$.
Here, we follow recommendations} given in Ref.~\cite{Schulze2008}. \coriii{The \gls{ps} properties are taken from the powder manufacturer data sheet~\cite{DynoseedsMSDS}.}
The powder density is defined as $\rho$*~$= \rho_{b} \, \varphi$, \gls{ps} density in bulk times the powder packing fraction.}
    \label{tab:Janssen_numparameters}
    \small{
      \renewcommand{\arraystretch}{1.05} 
      \begin{tabular}{l l c c c}
        \toprule
        {} & {\textbf{Parameter}}      & \textbf{Unit}               & \textbf{Value}  & \textbf{Absolute uncertainty}\\ \midrule
        {$\varphi$} & {packing fraction (initial) }             & {--}   & {0.55}	& --\\ 
        {$D$} & {container's diameter}        & {\si{\meter}} & {$5.40 \cdot 10^{-2}$}	& $10^{-4}$\\
        {$d$} & {piston's diameter}        & {\si{\meter}} & {$3.60 \cdot 10^{-2}$}	& $10^{-4}$\\
        {$m$} & {mass of powder}        & {\si{\kg}} & {$5.00 \cdot 10^{-2}$}	& $10^{-4}$ \\
        {$\rho_{b}$} & {density in bulk }             & {\si{\kilogram\per\meter\cubed}}   & {1050}	& 10	\\ 
        {$\rho$*} & {powder density }             & {\si{\kilogram\per\meter\cubed}}   & {580}	& 10	\\ 
        {$g$} & {gravitational constant }    & {\si{\meter\per\second\squared}}          & {9.81}	&-- \\
        {$\mu$} & {static friction coefficient }           & {--} & {0.3}	& --\\ 
        {$k$} & {Janssen coefficient }        & {--} & {0.4}	& {--} \\ 
        \bottomrule
    \end{tabular}}
\end{table*}

For the \acrlong{mug} experiments, $p_0$ is deducted from the steady-state
pressure (once a relatively stable value is reached after
transient decrease): under reduced gravitational acceleration and
before the start of piston rise (i.e.~at $t=\SI{0}{\s}$), the expected
pressure is $p(t=\SI{0}{\s}) = \SI{0}{\pascal}$, as no pressure is
applied on the sensors. Therefore, we assume that the remaining
pressure measured equals the offset $p_{0}$, and use this value to
estimate $p= p_{\text{exp}} - p_{0}$.

For the measurements conducted \acrlong{1g}, the reference value used is the
pressure at the beginning of the experiment, before the piston starts
rising. At this point, the pressure recorded should be solely due to
the weight of the powder inside the container: using the mass of the
powder at the sensor's height, we calculate the expected horizontal
pressure $p_h\equiv\Sigma_{rr}(h)$ from Eq.~(\ref{eq:6}). Numerical
parameters used are in Tab.~\ref{tab:Janssen_numparameters}.  The
offset pressure $p_0=p_{\text{exp}} - p_h$ is calculated for each
experiment to obtain $p$.

\subsection*{Formal uncertainty}
Formal uncertainty due to measurement limitations is calculated for
the packing fraction
from the mass $m$ of granulate, 
the density $\rho_b$ of the particles' material (in bulk),
and the volume occupied by the granular material, which depends
on the diameter $D$ squared and height $h$ of the piston.
As each variable is measured independently,
we rely on propagation of uncertainty laws
and calculate
the relative error $\delta \varphi / \varphi$ 
as the sum of relative errors on each variable.

The absolute uncertainties for each measured variable are given in Tab.~\ref{tab:Janssen_numparameters}.
The height $h$ of the piston is 
measured from image analysis using a 
scale placed right behind the piston's axis,
which results in an absolute uncertainty of 
$\delta h = \SI{0.5}{\mm}$.

The resulting absolute uncertainties $\delta \varphi$ are plotted as error bars in Fig.~\ref{fig:critical_height}.
The maximum absolute uncertainty
$\delta \varphi^{\text{gnd}, \, \textrm{$\mu$-g}} = 0.027$,
which represents a relative error of 5\%,
is taken as accuracy measurement for the average $\varphi^{\text{gnd}}_J $ and $\varphi^{\textrm{$\mu$-g}}_J $.

\subsection*{Statistical significance}
To validate the claim that the gravitational environment influences the packing fraction at jamming 
measured in our system,
a statistical significance test is essential. The test is done on the two series of four values 
for $\varphi^{\text{gnd}}_J $ and $\varphi^{\textrm{$\mu$-g}}_J $ given in Fig.~\ref{fig:critical_height}. 
The null-hypothesis is that the cumulative distributions are identical, 
$P(\varphi^{\text{gnd}}_J < \varphi_J) = P(\varphi^{\textrm{$\mu$-g}}_J < \varphi_J)$; 
the alternative hypothesis is that 
$P(\varphi^{\text{gnd}}_J < \varphi_J) < P(\varphi^{\textrm{$\mu$-g}}_J < \varphi_J)$.

All values given for packing fraction at jamming are calculated
solely from the experiment conducted 
on the first parabola of each flight,
before any powder has flowed under the piston (region \textnumero1 in Fig.~\ref{fig:regions_pistonbox}),
as shown in Fig.~\ref{fig:piston_probing_densification}(a--e).
Therefore, only four repetitions of the experiment are available.
For small samples where normal distribution cannot be assumed,
a nonparametric statistical significance test is preferable.
We use the
sum of ranks test known as
Mann-Whitney U-test~\cite{MannWhitney1947}.

We find a probability of the null-hypothesis of only 1.4\%~\cite{MannWhitney1947}.
We can therefore confidently conclude that for the limited amount of data available, 
the probability of
the distributions of $\varphi^{\text{gnd}}_J $ and $\varphi^{\textrm{$\mu$-g}}_J$
being equal is low enough to be rejected, 
which supports our conclusion that 
the packing densities at jamming are different between ground and microgravity.

\section*{Data Availability statement}

The datasets used in this article are publicly available in the Zenodo repository 7101542:
\url{https://doi.org/10.5281/zenodo.7101542}.

\section*{Code Availability statement}
The codes used to generate the plots are available from the corresponding author upon reasonable request.

\section*{Acknowledgments}

ODA acknowledges financial support of the \acrshort{esa}
\acrshort{npi} contract 4000122340 on \enquote{Physical Properties of
Powder-Based \acrshort{3d}-Printing in Space and On-Ground}, and of
the \acrshort{dlr}/\acrshort{daad} Research Fellowship 91647576.  WTK
acknowledges funding by the \acrshort{dfg} through grant number
KR4867/2. ODA and AH wish to express sincere gratitude to the
Spaceship \acrshort{eac} program for financial support.
Miranda Fateri is thanked for support during the \acrlong{pfc}.
Thorsten P\"oschel, Philip Born, Matthias Schr\"oter and Angel Santarossa are acknowledged for their critical reviewing of the manuscript.
Many thanks to the entire Novespace team for their help and patience,
and in particular to Yannick Bailh\'{e}.

\section*{Author contributions}

ODA and AH designed and build the setup and performed the measurements. ODA and WTK analyzed and interpreted the data and wrote the manuscript. AC and MS supported the project. MS procured the flight opportunity. All authors revised the manuscript.

\section*{Competing interests}
The authors declare no competing interests.

\section*{Additional information}
\textbf{Supplementary information} accompanies this paper.

\bibliographystyle{unsrtnat}
\bibliography{bibliography.bib}

\begin{thebibliography}{65}
\providecommand{\natexlab}[1]{#1}
\providecommand{\url}[1]{\texttt{#1}}
\expandafter\ifx\csname urlstyle\endcsname\relax
  \providecommand{\doi}[1]{doi: #1}\else
  \providecommand{\doi}{doi: \begingroup \urlstyle{rm}\Url}\fi

\bibitem[Wilkinson et~al.(2005)Wilkinson, Behringer, Jenkins, and
  Louge]{Wilkinson2005}
R.~A. Wilkinson, R.~P. Behringer, J.~T. Jenkins, and M.~Y. Louge.
\newblock Granular materials and the risks they pose for success on the moon
  and mars.
\newblock \emph{{AIP} Conference Proceedings}, 746:\penalty0 1216--1223, 2005.
\newblock \doi{10.1063/1.1867248}.

\bibitem[Karapiperis et~al.(2020)Karapiperis, Marshall, and
  Andrade]{Karapiperis2020}
K.~Karapiperis, J.~P. Marshall, and J.~E. Andrade.
\newblock Reduced gravity effects on the strength of granular matter: {DEM}
  simulations versus experiments.
\newblock \emph{Journal of Geotechnical and Geoenvironmental Engineering},
  146:\penalty0 06020005, 2020.
\newblock \doi{10.1061/(ASCE)GT.1943-5606.0002232}.

\bibitem[Antony et~al.(2021)Antony, Arowosola, Richter, Amanbayev, Barakat, and
  Pullithadathil]{Antony2021}
S.~J. Antony, B.~Arowosola, L.~Richter, T.~Amanbayev, T.~Barakat, and
  B.~Pullithadathil.
\newblock \emph{Modeling the flow characteristics of granular materials under
  low gravity environments using discrete element method}, pages 12--21.
\newblock 2021.
\newblock \doi{10.1061/9780784483374.002}.

\bibitem[D’Angelo et~al.(2021)D’Angelo, Kuthe, Liu, Wiedey, Bennett,
  Meisnar, Barnes, Kranz, Voigtmann, and Meyer]{dangelo2021}
O.~D’Angelo, F.~Kuthe, S.-J. Liu, R.~Wiedey, J.~M. Bennett, M.~Meisnar,
  A.~Barnes, W~T. Kranz, Th. Voigtmann, and A.~Meyer.
\newblock A gravity-independent powder-based additive manufacturing process
  tailored for space applications.
\newblock \emph{Additive Manufacturing}, 47:\penalty0 102349, 2021.

\bibitem[Mo et~al.(2021)Mo, Zhou, Gao, and Li]{Mo2021}
P.~Mo, G.~Zhou, F.~Gao, and R.~Li.
\newblock Bearing capacity of surface circular footings on granular material
  under low gravity fields.
\newblock \emph{Journal of Rock Mechanics and Geotechnical Engineering},
  13:\penalty0 612--625, 2021.
\newblock \doi{10.1016/j.jrmge.2020.11.009}.

\bibitem[de~Gennes(1999)]{deGennes1999}
P.~G. de~Gennes.
\newblock Granular matter: a tentative view.
\newblock \emph{Reviews of Modern Physics}, 71:\penalty0 S374--S382, 1999.
\newblock \doi{10.1103/RevModPhys.71.S374}.

\bibitem[Frenkel(2002)]{Frenkel2002}
D.~Frenkel.
\newblock Soft condensed matter.
\newblock \emph{Physica A}, 313:\penalty0 1--31, 2002.
\newblock \doi{10.1016/S0378-4371(02)01032-4}.

\bibitem[Karmali and Shelhamer(2008)]{Karmali2008}
F.~Karmali and M.~Shelhamer.
\newblock The dynamics of parabolic flight: flight characteristics and
  passenger percepts.
\newblock \emph{Acta Astronautica}, 63\penalty0 (5-6):\penalty0 594--602, 2008.
\newblock \doi{10.1016/j.actaastro.2008.04.009}.

\bibitem[Blum et~al.(2000)Blum, Wurm, Kempf, Poppe, Klahr, Kozasa, Rott,
  Henning, Dorschner, Schr{\"{a}}pler, Keller, Markiewicz, Mann, Gustafson,
  Giovane, Neuhaus, Fechtig, Gr{\"{u}}n, Feuerbacher, Kochan, Ratke, {El
  Goresy}, Morfill, Weidenschilling, Schwehm, Metzler, and Ip]{Blum2000}
J.~Blum, G.~Wurm, S.~Kempf, T.~Poppe, H.~Klahr, T.~Kozasa, M.~Rott, T.~Henning,
  J.~Dorschner, R.~Schr{\"{a}}pler, H.~U. Keller, W.~J. Markiewicz, I.~Mann,
  B.~A.S. Gustafson, F.~Giovane, D.~Neuhaus, H.~Fechtig, E.~Gr{\"{u}}n,
  B.~Feuerbacher, H.~Kochan, L.~Ratke, A.~{El Goresy}, G.~Morfill, S.~J.
  Weidenschilling, G.~Schwehm, K.~Metzler, and W.-H. Ip.
\newblock Growth and form of planetary seedlings: Results from a microgravity
  aggregation experiment.
\newblock \emph{Physical Review Letters}, 85:\penalty0 2426--2429, 2000.
\newblock \doi{10.1103/PhysRevLett.85.2426}.

\bibitem[Blum(2006)]{Blum2006}
J.~Blum.
\newblock Dust agglomeration.
\newblock \emph{Advances in Physics}, 55:\penalty0 881--947, 2006.
\newblock \doi{10.1080/00018730601095039}.

\bibitem[Weidling et~al.(2012)Weidling, G{\"{u}}ttler, and Blum]{Weidling2012}
R.~Weidling, C.~G{\"{u}}ttler, and J.~Blum.
\newblock Free collisions in a microgravity many-particle experiment. {I. Dust}
  aggregate sticking at low velocities.
\newblock \emph{Icarus}, 218:\penalty0 688--700, 2012.
\newblock \doi{10.1016/j.icarus.2011.10.002}.

\bibitem[Love et~al.(2014)Love, Petit, and Messenger]{Love2014}
S.~G. Love, D.~R. Petit, and S.~R. Messenger.
\newblock Particle aggregation in microgravity: Informal experiments on the
  international space station.
\newblock \emph{Meteoritics \& Planetary Science}, 49:\penalty0 732--739, 2014.
\newblock \doi{10.1111/maps.12286}.

\bibitem[Evesque(2004)]{Evesque2004}
P.~Evesque.
\newblock Distribution of contact forces in a homogeneous granular material of
  identical spheres under triaxial compression.
\newblock \emph{{Poudres {\&} Grains}}, 14:\penalty0 4, 2004.

\bibitem[Drescher and {de Jong de Josselin}(1972)]{Drescher1972}
A.~Drescher and G.~{de Jong de Josselin}.
\newblock Photoelastic verification of a mechanical model for the flow of a
  granular material.
\newblock \emph{Journal of the Mechanics and Physics of Solids}, 20:\penalty0
  337--351, 1972.
\newblock \doi{10.1016/0022-5096(72)90029-4}.

\bibitem[Liu et~al.(1995)Liu, Nagel, Schecter, Coppersmith, Majumdar, Narayan,
  and Witten]{Liu1995}
C.~H. Liu, S.~R. Nagel, D.~A. Schecter, S.~N. Coppersmith, S.~Majumdar,
  O.~Narayan, and T.~A. Witten.
\newblock Force fluctuations in bead packs.
\newblock \emph{Science}, 269:\penalty0 513--515, 1995.
\newblock ISSN 00368075.
\newblock \doi{10.1126/science.269.5223.513}.

\bibitem[Liu and Nagel(2001)]{Liu2001}
A.~J. Liu and S.~R. Nagel, editors.
\newblock \emph{Jamming and Rheology: Constrained Dynamics on Microscopic and
  Macroscopic Scales}.
\newblock Taylor {\&} Francis, New York, 2001.
\newblock ISBN 978-0748408795.

\bibitem[Liu and Nagel(1998)]{Liu1998}
A.~J. Liu and S.~R. Nagel.
\newblock {Jamming is not just cool any more}.
\newblock \emph{Nature}, 396:\penalty0 21--22, 1998.
\newblock \doi{https://doi.org/10.1038/23819}.

\bibitem[Brown and Jaeger(2014)]{Brown2014}
E.~Brown and H.~M. Jaeger.
\newblock Shear thickening in concentrated suspensions: phenomenology,
  mechanisms and relations to jamming.
\newblock \emph{Reports on progress in physics}, 77:\penalty0 4, 2014.
\newblock \doi{10.1088/0034-4885/77/4/046602}.

\bibitem[Majmudar et~al.(2007)Majmudar, Sperl, Luding, and
  Behringer]{Majmudar2007}
T.~S. Majmudar, M.~Sperl, S.~Luding, and R.~P. Behringer.
\newblock Jamming transition in granular systems.
\newblock \emph{Physical Review Letters}, 98:\penalty0 058001, 2007.
\newblock \doi{10.1103/PhysRevLett.98.058001}.

\bibitem[Behringer and Chakraborty(2018)]{Behringer2018}
R.~P. Behringer and B.~Chakraborty.
\newblock The physics of jamming for granular materials: a review.
\newblock \emph{Reports on Progress in Physics}, 82:\penalty0 012601, 2018.
\newblock \doi{10.1088/1361-6633/aadc3c}.

\bibitem[Cates et~al.(1999)Cates, Wittmer, Bouchaud, and Claudin]{Cates1999}
M.~E. Cates, J.~P. Wittmer, J.~P. Bouchaud, and P.~Claudin.
\newblock Jamming and stress propagation in particulate matter.
\newblock \emph{Physica A}, 263:\penalty0 354--361, 1999.
\newblock \doi{10.1016/s0378-4371(98)00491-9}.

\bibitem[Bi et~al.(2011)Bi, Zhang, Chakraborty, and Behringer]{Bi2011}
D.~Bi, J.~Zhang, B.~Chakraborty, and R.~P. Behringer.
\newblock Jamming by shear.
\newblock \emph{Nature}, 480:\penalty0 7377, 2011.
\newblock \doi{10.1038/nature10667}.

\bibitem[Wang et~al.(2013)Wang, Zhu, Luding, and Yu]{Wang2013}
X.~Wang, H.~P. Zhu, S.~Luding, and A.~B. Yu.
\newblock Regime transitions of granular flow in a shear cell: A
  micromechanical study.
\newblock \emph{Physical Review E}, 88:\penalty0 032203, 2013.
\newblock \doi{10.1103/PhysRevE.88.032203}.

\bibitem[Grob et~al.(2014)Grob, Heussinger, and Zippelius]{Grob2014}
M.~Grob, C.~Heussinger, and A.~Zippelius.
\newblock Jamming of frictional particles: A nonequilibrium first-order phase
  transition.
\newblock \emph{Physical Review E}, 89:\penalty0 050201, 2014.
\newblock \doi{10.1103/PhysRevE.89.050201}.

\bibitem[Onoda and Liniger(1990)]{Onoda1990}
G.~Y. Onoda and E.~G. Liniger.
\newblock Random loose packings of uniform spheres and the dilatancy onset.
\newblock \emph{Physical Review Letters}, 64:\penalty0 2727--2730, 1990.
\newblock \doi{10.1103/PhysRevLett.64.2727}.

\bibitem[M{\'{e}}tayer et~al.(2011)M{\'{e}}tayer, III, Radin, Swinney, and
  Schr\"{o}ter]{Metayer2011}
J.-F. M{\'{e}}tayer, D.~J.~Suntrup III, C.~Radin, H.~L. Swinney, and
  M.~Schr\"{o}ter.
\newblock Shearing of frictional sphere packings.
\newblock \emph{{EPL} (Europhysics Letters)}, 93\penalty0 (6):\penalty0 64003,
  2011.
\newblock \doi{10.1209/0295-5075/93/64003}.

\bibitem[Song et~al.(2008)Song, Wang, and Makse]{Song2008}
C.~Song, P.~Wang, and H.~A. Makse.
\newblock A phase diagram for jammed matter.
\newblock \emph{Nature}, 453:\penalty0 7195, 2008.
\newblock \doi{10.1038/nature06981}.

\bibitem[Torquato et~al.(2000)Torquato, Truskett, and
  Debenedetti]{Torquato2000}
S.~Torquato, T.~M. Truskett, and P.~G. Debenedetti.
\newblock Is random close packing of spheres well defined?
\newblock \emph{Physical Review Letters}, 84:\penalty0 2064--2067, 2000.
\newblock \doi{10.1103/PhysRevLett.84.2064}.

\bibitem[Kumar and Luding(2016)]{Kumar2016}
N.~Kumar and S.~Luding.
\newblock Memory of jamming--multiscale models for soft and granular matter.
\newblock \emph{Granular Matter}, 18:\penalty0 3, 2016.
\newblock \doi{10.1007/s10035-016-0624-2}.

\bibitem[Chaudhuri et~al.(2010)Chaudhuri, Berthier, and Sastry]{Chaudhuri2010}
P.~Chaudhuri, L.~Berthier, and S.~Sastry.
\newblock Jamming transitions in amorphous packings of frictionless spheres
  occur over a continuous range of volume fractions.
\newblock \emph{Physical Review Letters}, 104:\penalty0 165701, 2010.
\newblock \doi{10.1103/PhysRevLett.104.165701}.

\bibitem[Torquato and Stillinger(2010)]{Torquato2010}
S.~Torquato and F.~H. Stillinger.
\newblock Jammed hard-particle packings: From kepler to bernal and beyond.
\newblock \emph{Rev. Mod. Phys.}, 82:\penalty0 2633--2672, 2010.
\newblock \doi{10.1103/RevModPhys.82.2633}.

\bibitem[van Hecke(2009)]{VanHecke2009}
M.~van Hecke.
\newblock Jamming of soft particles: geometry, mechanics, scaling and
  isostaticity.
\newblock \emph{Journal of Physics: Condensed Matter}, 22:\penalty0 033101,
  2009.
\newblock \doi{10.1088/0953-8984/22/3/033101}.

\bibitem[Hermes and Dijkstra(2010)]{Hermes2010}
M.~Hermes and M.~Dijkstra.
\newblock Jamming of polydisperse hard spheres: the effect of kinetic arrest.
\newblock \emph{{EPL} (Europhysics Letters)}, 89:\penalty0 38005, 2010.
\newblock \doi{10.1209/0295-5075/89/38005}.

\bibitem[Pica~Ciamarra et~al.(2010)Pica~Ciamarra, Coniglio, and
  de~Candia]{Ciamarra2010}
M.~Pica~Ciamarra, A.~Coniglio, and A.~de~Candia.
\newblock Disordered jammed packings of frictionless spheres.
\newblock \emph{Soft Matter}, 6:\penalty0 13, 2010.
\newblock \doi{10.1039/C001904F}.

\bibitem[Otsuki and Hayakawa(2012)]{Otsuki2012}
M.~Otsuki and H.~Hayakawa.
\newblock Critical scaling of a jammed system after a quench of temperature.
\newblock \emph{Physical Review E}, 86:\penalty0 031505, 2012.
\newblock \doi{10.1103/PhysRevE.86.031505}.

\bibitem[Baranau and Tallarek(2014)]{Baranau2014}
V.~Baranau and U.~Tallarek.
\newblock Random-close packing limits for monodisperse and polydisperse hard
  spheres.
\newblock \emph{Soft Matter}, 10:\penalty0 21, 2014.
\newblock \doi{10.1039/C3SM52959B}.

\bibitem[Ciamarra et~al.(2011)Ciamarra, Pastore, Nicodemi, and
  Coniglio]{Ciamarra2011}
M.~P. Ciamarra, R.~Pastore, M.~Nicodemi, and A.~Coniglio.
\newblock Jamming phase diagram for frictional particles.
\newblock \emph{Physical Review E}, 84:\penalty0 041308, 2011.
\newblock \doi{10.1103/PhysRevE.84.041308}.

\bibitem[Mari et~al.(2014)Mari, Seto, Morris, and Denn]{Mari2014}
R.~Mari, R.~Seto, J.~F. Morris, and M.~M. Denn.
\newblock {Shear thickening, frictionless and frictional rheologies in
  non-Brownian suspensions}.
\newblock \emph{Journal of Rheology}, 58:\penalty0 1693, 2014.
\newblock \doi{10.1122/1.4890747}.

\bibitem[Jerkins et~al.(2008)Jerkins, Schr\"oter, Swinney, Senden, Saadatfar,
  and Aste]{Jerkins2008}
M.~Jerkins, M.~Schr\"oter, H.~L. Swinney, T.~J. Senden, M.~Saadatfar, and
  T.~Aste.
\newblock Onset of mechanical stability in random packings of frictional
  spheres.
\newblock \emph{Physical Review Letters}, 101:\penalty0 018301, 2008.
\newblock \doi{10.1103/PhysRevLett.101.018301}.

\bibitem[Silbert(2010)]{Silbert2010}
L.~E. Silbert.
\newblock Jamming of frictional spheres and random loose packing.
\newblock \emph{Soft Matter}, 6:\penalty0 2918--2924, 2010.
\newblock \doi{10.1039/C001973A}.

\bibitem[Farrell et~al.(2010)Farrell, Martini, and Menon]{Farrell2010}
G.~Farrell, K.~M. Martini, and N.~Menon.
\newblock Loose packings of frictional spheres.
\newblock \emph{Soft Matter}, 6:\penalty0 2925--2930, 2010.

\bibitem[Noirhomme et~al.(2017)Noirhomme, Ludewig, Vandewalle, and
  Opsomer]{Noirhomme2017}
M.~Noirhomme, F.~Ludewig, N.~Vandewalle, and E.~Opsomer.
\newblock Cluster growth in driven granular gases.
\newblock \emph{Physical Review E}, 95:\penalty0 022905, 2017.
\newblock \doi{10.1103/PhysRevE.95.022905}.

\bibitem[Abed~Zadeh et~al.(2019)Abed~Zadeh, Bar{\'e}s, Brzinski, Daniels,
  Dijksman, Docquier, Everitt, Kollmer, Lantsoght, Wang, Workamp, Zhao, and
  Zheng]{Zadeh2019}
A.~Abed~Zadeh, J.~Bar{\'e}s, T.~A. Brzinski, K.~E. Daniels, J.~Dijksman,
  N.~Docquier, H.~O. Everitt, J.~E. Kollmer, O.~Lantsoght, D.~Wang, M.~Workamp,
  Y.~Zhao, and H.~Zheng.
\newblock Enlightening force chains: a review of photoelasticimetry in granular
  matter.
\newblock \emph{Granular Matter}, 21:\penalty0 83, 2019.
\newblock \doi{10.1007/s10035-019-0942-2}.

\bibitem[Daniels et~al.(2017)Daniels, Kollmer, and Puckett]{Daniels2017}
K.~E. Daniels, J.~E. Kollmer, and J.~G. Puckett.
\newblock Photoelastic force measurements in granular materials.
\newblock \emph{Review of Scientific Instruments}, 88\penalty0 (5):\penalty0
  051808, 2017.
\newblock \doi{10.1063/1.4983049}.

\bibitem[Carr et~al.(2018)Carr, Bryan, Saboda, Bhattaru, Ruvkun, and
  Zuber]{Carr2018}
C.~E. Carr, N.~C. Bryan, K.~N. Saboda, S.~A. Bhattaru, G.~Ruvkun, and M.~T.
  Zuber.
\newblock Acceleration profiles and processing methods for parabolic flight.
\newblock \emph{npj Microgravity}, 4, 2018.
\newblock \doi{10.1038/s41526-018-0050-3}.

\bibitem[Makse et~al.(2005)Makse, Bruji{\'c}, and Edwards]{Makse2004}
H.~A. Makse, J.~Bruji{\'c}, and S.~F. Edwards.
\newblock \emph{Statistical Mechanics of Jammed Matter}, chapter~3, pages
  45--85.
\newblock John Wiley \& Sons, Ltd, 2005.
\newblock ISBN 9783527603626.
\newblock \doi{10.1002/352760362x.ch3}.

\bibitem[O'Hern et~al.(2002)O'Hern, Langer, Liu, and Nagel]{OHern2002}
C.~S. O'Hern, S.~A. Langer, A.~J. Liu, and S.~R. Nagel.
\newblock Random packings of frictionless particles.
\newblock \emph{Physical Review Letters}, 88:\penalty0 075507, 2002.
\newblock \doi{10.1103/PhysRevLett.88.075507}.

\bibitem[Farr et~al.(1997)Farr, Melrose, and Ball]{Farr1997}
R.~S. Farr, J.~R. Melrose, and R.~C. Ball.
\newblock Kinetic theory of jamming in hard-sphere startup flows.
\newblock \emph{Physical Review E}, 55:\penalty0 7203--7211, 1997.
\newblock \doi{10.1103/PhysRevE.55.7203}.

\bibitem[Murdoch et~al.(2013)Murdoch, Rozitis, Nordstrom, Green, Michel, {De
  Lophem}, and Losert]{Murdoch2013}
N.~Murdoch, B.~Rozitis, K.~Nordstrom, S.~F. Green, P.~Michel, T.~L. {De
  Lophem}, and W.~Losert.
\newblock Granular convection in microgravity.
\newblock \emph{Phy. Rev. Lett.}, 110:\penalty0 018307, 2013.
\newblock \doi{10.1103/PhysRevLett.110.018307}.

\bibitem[Janssen(1895)]{Janssen1895}
H.~A. Janssen.
\newblock {Versuche {\"{u}}ber Getreidedruck in Silozellen} [{On} grain
  pressure in silos].
\newblock \emph{Zeitschrift des Vereins Deutscher Ingenieure}, 39:\penalty0 35,
  1895.

\bibitem[Sollich et~al.(1997)Sollich, Lequeux, H{\'e}braud, and
  Cates]{Sollich1997}
P.~Sollich, F.~Lequeux, P.~H{\'e}braud, and M.~E. Cates.
\newblock Rheology of soft glassy materials.
\newblock \emph{Physical Review Letters}, 78:\penalty0 2020, 1997.

\bibitem[Fielding et~al.(2000)Fielding, Sollich, and Cates]{Fielding2000}
S.~M Fielding, P.~Sollich, and M.~E. Cates.
\newblock Aging and rheology in soft materials.
\newblock \emph{Journal of Rheology}, 44:\penalty0 2, 2000.

\bibitem[Bi and Chakraborty(2009)]{Bi2009}
D.~Bi and B.~Chakraborty.
\newblock Rheology of granular materials: dynamics in a stress landscape.
\newblock \emph{Phil. Trans. Royal Soc. A}, 367:\penalty0 1909, 2009.

\bibitem[Blumenfeld and Edwards(2009)]{Blumenfeld2009}
R.~Blumenfeld and S.~F. Edwards.
\newblock On granular stress statistics: Compactivity, angoricity, and some
  open issues.
\newblock \emph{J. Phys. Chem. B}, 113:\penalty0 12, 2009.

\bibitem[Nedderman(1992)]{Nedderman1992}
R.~M. Nedderman.
\newblock \emph{{Statistics and Kinematics of Granular Materials}}.
\newblock Cambridge University Press, 1992.
\newblock ISBN 0-521-40435-5.

\bibitem[Wyart and Cates(2014)]{Wyart2014}
M.~Wyart and M.~E. Cates.
\newblock Discontinuous shear thickening without inertia in dense
  non-{Brownian} suspensions.
\newblock \emph{Physical Review Letters}, 112:\penalty0 098302, 2014.
\newblock \doi{10.1103/PhysRevLett.112.098302}.

\bibitem[Sture et~al.(1998)Sture, Costes, Batiste, Lankton, Alshibli, Jeremic,
  Swanson, and Frank]{Sture.1998}
S.~Sture, N.~C. Costes, S.~N. Batiste, M.~R. Lankton, K.~A. Alshibli,
  B.~Jeremic, R.~A. Swanson, and M.~Frank.
\newblock Mechanics of granular materials at low effective stresses.
\newblock \emph{Journal of Aerospace Engineering}, 11:\penalty0 67--72, 1998.

\bibitem[Alshibli et~al.(2003)Alshibli, Batiste, and Sture]{Alshibli2003}
K.~A. Alshibli, S.~N. Batiste, and S.~Sture.
\newblock Strain localization in sand: Plane strain versus triaxial
  compression.
\newblock \emph{Journal of Geotechnical and Geoenvironmental Engineering},
  129\penalty0 (6):\penalty0 483--494, 2003.
\newblock \doi{10.1061/(ASCE)1090-0241(2003)129:6(483)}.

\bibitem[AS(2010)]{DynoseedsMSDS}
Microbeads AS.
\newblock Safety data sheet dynoseeds ts.
\newblock
  \url{http://www.micro-beads.com/UserFiles/file/pdf%202010/MSDS%20Dynoseeds%20TS,%202010.pdf},
  2010.
\newblock [Online] Accessed: 2022-06-08.

\bibitem[Shaebani et~al.(2012)Shaebani, Madadi, Luding, and Wolf]{Shaebani2012}
M.~R. Shaebani, M.~Madadi, S.~Luding, and D.~E. Wolf.
\newblock Influence of polydispersity on micromechanics of granular materials.
\newblock \emph{Physical Review E}, 85:\penalty0 011301, 2012.
\newblock \doi{10.1103/PhysRevE.85.011301}.

\bibitem[Hertz(1882)]{Hertz1882}
H.~Hertz.
\newblock {\"Uber die Ber\"uhrung fester elastischer K\"orper} {[About the
  contact of solid elastic bodies]}.
\newblock \emph{Journal f\"ur die Reine und Angewandte Mathematik},
  92:\penalty0 156--171, 1882.

\bibitem[Wypych(2016)]{Wypych2016}
G.~Wypych.
\newblock {PS} (polystyrene).
\newblock In \emph{Handbook of Polymers}, pages 560--566. ChemTec Publishing,
  second edition, 2016.
\newblock \doi{10.1016/B978-1-895198-92-8.50175-0}.

\bibitem[Mavko et~al.(2009)Mavko, Mukerji, and Dvorkin]{mavko2009}
Gary Mavko, Tapan Mukerji, and Jack Dvorkin.
\newblock \emph{The rock physics handbook}.
\newblock Cambridge University Press, Cambridge, 2nd edition, 2009.

\bibitem[Schulze(2008)]{Schulze2008}
D.~Schulze.
\newblock \emph{Powders and Bulk Solids -- Behavior, Characterization, Storage
  and Flow}.
\newblock Springer-Verlag Berlin Heidelberg, Berlin, 2008.
\newblock \doi{10.1007/978-3-540-73768-1}.

\bibitem[Mann and Whitney(1947)]{MannWhitney1947}
H.~B. Mann and D.~R. Whitney.
\newblock {On a Test of Whether one of Two Random Variables is Stochastically
  Larger than the Other}.
\newblock \emph{The Annals of Mathematical Statistics}, 18:\penalty0 50--60,
  1947.
\newblock \doi{10.1214/aoms/1177730491}.

\end{thebibliography}


\end{document}